\documentclass[a4paper,fleqn,usenatbib]{mn2e}

\usepackage{aas_macros}
\usepackage{color}
\usepackage{bm}
\usepackage[draft]{hyperref}
\usepackage{newtxtext,newtxmath}
\usepackage[T1]{fontenc}
\usepackage{times}
\usepackage{amssymb}
\usepackage{float}
\usepackage{graphicx}
\usepackage{epstopdf}
\usepackage{lscape}
\usepackage{threeparttable}
\usepackage{bigints}

\def\kms{km s$^{-1}$ }
\newcommand{\Gaia}{{\it Gaia} }

\title[The fastest stars in \Gaia DR2]{\Gaia DR2 in 6D: Searching for the fastest stars in the Galaxy}

\author[T. Marchetti et al.]{T. Marchetti$^{1}$\thanks{E-mail: marchetti@strw.leidenuniv.nl}, E. M. Rossi$^{1}$ and A. G. A. Brown$^{1}$\\
$^1$Leiden Observatory, Leiden University, PO Box 9513 2300 RA Leiden, the Netherlands \\
}

\begin{document}

\pagerange{\pageref{firstpage}--\pageref{lastpage}} \pubyear{2017}

\maketitle

\label{firstpage}

\begin{abstract}

We search for the fastest stars in the subset of stars with radial velocity measurements of the second data release (DR2) of the European Space Agency mission \Gaia. Starting from the observed positions, parallaxes, proper motions, and radial velocities, we construct the distance and total velocity distribution of more than $7$ million stars in our Milky Way, deriving the full 6D phase space information in Galactocentric coordinates. These information are shared in a catalogue, publicly available at \url{http://home.strw.leidenuniv.nl/~marchetti/research.html}. To search for unbound stars, we then focus on stars with a probability greater than $50 \%$ of being unbound from the Milky Way. This cut results in a clean sample of $125$ sources with reliable astrometric parameters and radial velocities. Of these, $20$ stars have probabilities greater than 80 $\%$ of being unbound from the Galaxy. On this latter sub-sample, we perform orbit integration to characterize the stars' orbital parameter distributions. As expected given the relatively small sample size of bright stars, we find no hypervelocity star candidates, stars that are moving on orbits consistent with coming from the Galactic Centre. Instead, we find $7$ hyper-runaway star candidates, coming from the Galactic disk. Surprisingly, the remaining $13$ unbound stars cannot be traced back to the Galaxy, including two of the fastest stars (around $700$ \kms). If conformed, these may constitute the tip of the iceberg of a large extragalactic population or the extreme velocity tail of stellar streams.

\end{abstract}


\section{Introduction}
\label{sec:intro}

Stars with extremely high velocities have been long studied to probe our Galaxy. The interest in the high velocity tail of the total velocity distribution of stars in our Milky Way is twofold. First, it flags the presence of extreme dynamical and astrophysical processes, especially when the velocity of a star is so high that it approaches (or even exceeds) the escape speed from the Galaxy at its position. Secondly, high velocity stars, spanning a large range of distances, can be used as dynamical tracers of integral properties of the Galaxy. The stellar high velocity distribution has for example been used to trace the local Galactic escape speed and the mass of the Milky Way \citep[e.g.][]{smith+07, gnedin10, Piffl}. To put the concept of \emph{high velocity} in context, the value of the escape speed is found to be $\sim 530$ \kms at the Sun position, it increases up to $\sim 600$ \kms in the central regions of the Galaxy, and then falls down to $\lesssim 400$ \kms at Galactocentric distances $\sim 50$ kpc \citep{williams+17}.

A first class of objects that can be found in the high tail of the total velocity distribution is fast halo stars. Their measured dispersion velocity is around 150 \kms \citep{smith+09,evans+16}, therefore 3-$\sigma$ outliers can exceed $450$ \kms, while remaining bound. Halo stars could also reach unbound velocities, when they are part of the debris of tidally disrupted satellite galaxies, like the Sagittarius Dwarf galaxy, that has not yet virialized \citep[e.g.][]{abadi+09}. Velocities outliers in the bulge and disk velocity distribution may also exist and become apparent in a large data set.

"Runaway stars" (RSs) form an another class of high velocity stars. They were originally introduced as O and B type stars ejected from the Galactic disk with velocities higher than $40$ \kms \citep{blaauw61}. Theoretically, there are two main formation channels: i) dynamical encounters between stars in dense stellar systems such as young star clusters \citep[e.g.][]{poveda+67, leonard+90, gvaramadze+09}, and ii) supernova explosions in stellar binary systems \citep[e.g.][]{blaauw61, portegieszwart00}. Both mechanisms have been shown to occur in our Galaxy \citep{hoogerwerf+01}. Typical velocities attained by the two formation channels are of the order of a few tens of \kms, and even if several hundreds of \kms can be attained for the most extreme systems \citep{portegieszwart00, przybilla+08, gvaramadze+09, gvaramadze+11, silva+11}, simulations indicate that the majority of runaway stars from dynamical encounters have ejection velocities $\lesssim 200$ \kms \citep{perets12}. Recent results show that it is possible to achieve ejection velocities up to $\sim 1300$ \kms for low-mass G/K type stars in very compact binaries \citep{tauris15}. Nevertheless, the rate of production of unbound RSs, referred to as \emph{hyper runaway stars} (HRSs), is estimated to be as low as $8 \cdot 10^{-7}$ yr$^{-1}$ \citep{perets12, brown15}. 

As a class, the fastest stars in our Galaxy are expected to be hypervelocity stars (HVSs). These were first theoretically predicted by \cite{hills88} as the result of a three-body interaction between a binary star and the massive black hole in the Galactic Centre (GC), Sagittarius A$^*$. Following this close encounter, a star can be ejected with a velocity $\sim 1000$ \kms, sufficiently high to escape from the gravitational field of the Milky Way \citep{kenyon+08, brown15}. The first HVS candidate was discovered by \cite{brown+05}: a B-type star with a velocity more than twice the Galactic escape speed at its position. Currently about $\sim 20$ unbound HVSs with velocities $\sim 300$ - $700$ \kms have been discovered by targeting young stars in the outer halo of the Milky Way \citep{Brown+14}. In addition, tens of mostly bound candidates have been found at smaller distances but uncertainties prevent the precise identification of the GC as their ejection location \citep[e.g.][]{hawkins+15, vickers+15, zhang+16, marchetti+17, ziegerer+17}. HVSs are predicted to be ejected from the GC with an uncertain rate around $10^{-4}$ yr$^{-1}$ \citep{yu&tremaine03, zhang+13}, two orders of magnitude larger than the rate of ejection of runaway stars with comparable velocities from the stellar disk \citep{brown15}. Because of their extremely high velocities, HVS trajectories span a large range of distances, from the GC to the outer halo. Thus HVSs have been proposed as tools to study the matter distribution in our Galaxy \citep[e.g.][]{gnedin+05, sesana+07, kenyon+14, Rossi+17, fragione&loeb16, contigiani+18} and the GC environment \citep[e.g.][]{zhang+13, madigan+14}, but a larger and less observationally biased sample is needed in order to break degeneracies between the GC binary content and the Galactic potential parameters \citep{Rossi+17}. Using the fact that their angular momentum should be very close to zero, HVSs have also been proposed as tools to constrain the Solar position and velocity \citep{hattori+18}. Other possible alternative mechanisms leading to the acceleration of HVSs are the encounter between a single star and a massive black hole binary in the GC \citep[e.g.][]{yu&tremaine03, sesana+06, sesana+08}, the interaction between a globular cluster with a single or a binary massive black hole in the GC \citep{capuzzodolcetta+15, fragione+16}, and the tidal interaction of a dwarf galaxy near the center of the Galaxy \citep{abadi+09}. Another possible ejection origin for HVSs and high velocity stars in our Galaxy is the Large Magellanic Cloud \citep[LMC, ][]{boubert+16,boubert+17a,erkal+18}, orbiting the Milky Way with a velocity $\sim 380$ \kms \citep{vdMarel+14}. 

In addition to the unbound population of HVSs, all the ejection mechanisms mentioned above predict also a population of \emph{bound} HVSs (BHVSs): stars sharing the same formation scenario as HVSs, but with an ejection velocity which is not sufficiently high to escape from the whole Milky Way \citep[e.g.][]{bromley+06}. Most of the deceleration occurs in the inner few kpc due to the bulge potential \citep{kenyon+08}, and the minimum velocity necessary at ejection to be unbound is of the order of $\sim 800$ \kms \citep[a precise value depends on the choice of the Galactic potential,][]{brown15, Rossi+17}. If we consider the Hills mechanism , this population of bound stars is expected to be dominant over the sample of HVSs \citep{rossi+14, marchetti+18}.

At the moment, the fastest star discovered in our Galaxy is US 708, traveling away from the Milky Way with a total velocity $\sim 1200$ \kms \citep{hirsch+05}. Its orbit is not consistent with coming from the GC \citep{brown+15}, and the most likely mechanism responsible for its acceleration is the explosion of a thermonuclear supernova in an ultra-compact binary in the Galactic disk \citep{geier+15}.

The second data release (DR2) of the European Space Agency satellite \Gaia \citep{gaiaa, gaiadr2} gives us the first opportunity to look for extremely high velocity stars in our Milky Way, using an unprecedented sample of precisely and accurately measured sources. On 2018 April 25, \Gaia provided positions $(\alpha, \delta)$, parallaxes $\varpi$ and proper motions $(\mu_{\alpha*}, \mu_{\delta})$ for more than $1.3$ billion of stars, and, notably, radial velocities $v_\mathrm{rad}$ for a subset of $7224631$ stars brighter than the $12$th magnitude in the \Gaia Radial Velocity Spectrograph (RVS) passband \citep{cropper+18,katz+18}. Radial velocities are included in the \Gaia catalogue for stars with an effective temperature $T_\mathrm{eff}$ from $3550$ to $6990$ K, and have typical uncertainties of the order of few hundreds of m s$^{-1}$ at the bright end of the magnitude distribution (\Gaia $G$ band magnitude $\approx 4$), and of a few \kms at the faint end ($G \approx 13$). 

Using \Gaia DR2 data, \cite{boubert+18} show that almost all the previously discovered late-type HVS candidates are most likely bound to the Galaxy, and their total velocity was previously overestimated because of inaccurate parallaxes and/or proper motions. Only one late-type star, LAMOST J115209.12+120258.0 \citep{li+15}, is most likely unbound, but the Hills mechanisms is ruled out as a possible explanation of its extremely high velocity. The majority of B-type HVSs from \citep{Brown+14,brown+15} are still found to be consistent with coming from the GC when using \Gaia DR2 proper motions \citep{erkal+18}. 

In this paper we search for the fastest stars in the Milky Way, within the sample of $\sim 7$ million stars with a six-dimensional phase space measurement in \Gaia DR2. Since the origin of high velocity stars in our Galaxy is still a puzzling open question, we simply construct the total velocity distribution in the Galactic rest-frame in order to identify and characterize the high velocity tail. In doing so, we do not bias our search towards any specific class of high velocity stars. 

This manuscript is organized as follows. In Section \ref{sec:dist}, we explain how we determine distances and total velocities in the Galactic rest frame for the whole sample of stars. We presents results in terms of stellar total velocity in Section \ref{sec:v_distr}. In Section \ref{sec:highV}, we focus on the high velocity stars in the sample, and then in Section \ref{sec:highVcand} we concentrate on the stars with a probability greater than $80 \%$ of being unbound from the Galaxy, discussing individually the most interesting candidates. Finally, we conclude and discuss our results and findings in Section \ref{sec:discussion}.

\section{Distance and Total Velocity Determination}
\label{sec:dist}

The \Gaia catalogue provides parallaxes, and thus a conversion to a distance is required to convert the apparent motion of an object on the celestial sphere to a physical motion in space, that is needed to determine the total velocity of a star. \cite{bailer-jones} discusses in details how this operation is not trivial when the relative error in parallax, $f \equiv \sigma_\varpi / \varpi$, is either above $20\%$ or it is negative. We choose to separate the discussion on how we determine distances and total velocities of stars with $0 < f \le 0.1$ (the "\emph{low-f} sample") and of those with either $f > 0.1$ or $f < 0$ (the "\emph{high-f} sample").
There are $7183262$ stars with both radial velocity and the astrometric parameters (parallax and proper motions) in \Gaia DR2, therefore in the following we will focus on this subsample of stars.

\subsection{The "\emph{low-f} Sample"}
\label{sec:lowf}

$5393495$ out of $7183262$ stars ($\sim 75 \%$) with radial velocity measurement in \Gaia DR2 have a relative error in parallax $0 < f \le 0.1$. For this majority of stars we can get an accurate determination of their distance just by inverting the parallax: $d = 1/\varpi$ \citep{bailer-jones}. We then model the proper motions and parallax distribution as a multivariate Gaussian with mean vector:
\begin{equation}
\label{eq:mean}
\mathbf{m} = [\mu_{\alpha*}, \mu_\delta, \varpi]
\end{equation}
and with covariance matrix:
\begin{equation}
\label{eq:covmatr} 
\footnotesize
\arraycolsep=2pt
\thickmuskip =1.mu
\Sigma = {}
\left(
\begin{array}{@{}ccc@{}}
\sigma_{\mu_{\alpha *}}^2 & \sigma_{\mu_{\alpha *}} \sigma_{\mu_\delta} \rho(\mu_{\alpha *},\mu_\delta) & \sigma_{\mu_{\alpha *}} \sigma_\varpi \rho(\mu_{\alpha *},\varpi) \\
\sigma_{\mu_{\alpha *}} \sigma_{\mu_\delta} \rho(\mu_{\alpha *},\mu_\delta) & \sigma_{\mu_\delta}^2 & \sigma_{\mu_{\delta}} \sigma_\varpi \rho(\mu_\delta,\varpi) \\
\sigma_{\mu_{\alpha *}} \sigma_\varpi \rho(\mu_{\alpha *},\varpi) & \sigma_{\mu_\delta} \sigma_\varpi \rho(\mu_\delta,\mu_\varpi) & \sigma_\varpi^2 
\end{array}
\right),
\end{equation}
where $\rho(i,j)$ denotes the correlation coefficient between the astrometric parameters $i$ and $j$, and it is provided in the \Gaia DR2 catalogue. Radial velocities are uncorrelated to the astrometric parameters, and we assume them to follow a Gaussian distribution centered on $v_\mathrm{rad}$, and with standard deviation $\sigma_{v_\mathrm{rad}}$. We then draw 1000 Monte Carlo (MC) realizations of each star's observed astrometric parameters, and we simply compute distances by inverting parallaxes. 

Total velocities in the Galactic rest frame are computed correcting radial velocities and proper motions for the solar and the local standard of rest (LSR) motion \citep{schonrich}. In doing so, we assume that the distance between the Sun and the GC is $d_\odot = 8.2$ kpc, and that the Sun has an height above the stellar disk of $z_\odot = 25$ pc \citep{bland-hawthorn+16}. We assume a rotation velocity at the Sun position $v_\mathrm{LSR} = 238$ \kms and a Sun's peculiar velocity vector $ \mathbf{v_\odot} = [U_\odot, V_\odot, W_\odot] = [14.0, 12.24, 7.25]$ \kms \citep{schonrich09,schonrich,bland-hawthorn+16}. To save computational time, we do not sample within the uncertainties of the Solar position and motion. We verify that this does not considerably affect our results. We then derive Galactic rectangular velocities $(U, V, W)$ adopting the following convention: $U$ is positive when pointing in the direction of the GC, $V$ is positive along the direction of the Sun rotation around the Galaxy, and $W$ is positive when pointing towards the North Galactic Pole \citep{johnson_soderblom}. Starting from the MC samples on proper motions, distances, and radial velocities, we then compute total velocities in the Galactic rest frame $v_\mathrm{GC} = v_\mathrm{GC}(\alpha, \delta, \mu_{\alpha*}, \mu_{\delta}, d, v_\mathrm{rad})$ summing in quadrature the three velocity components $(U, V, W)$.

Finally, for each star we estimate the probability $P_\mathrm{ub}$ of being unbound from the Galaxy as the fraction of MC realizations which result in a total velocity $v_\mathrm{GC}$ greater than the escape speed from the MW at that given position. We compute the escape velocity from the Galaxy at each position using the Galactic potential model introduced and discussed in Section \ref{sec:orbits}.

\subsection{The "\emph{high-f} Sample"}
\label{sec:highf}

A more careful analysis is required for $1789767$ stars ($\sim 25 \%$) with either $f > 0.1$ or with a negative measured parallax. For these stars, we follow the approach outlined in \cite{bailer-jones, astraatmadjaI, astraatmadjaII, luri+18, bailerjones+18}. We use a full Bayesian analysis to determine the posterior probability $P(d|\varpi, \sigma_\varpi)$ of observing a star at a distance $d$, given the measured parallax $\varpi$ and its Gaussian uncertainty $\sigma_\varpi$. The authors show how the choice of the prior probability on distance $P(d)$ can seriously affect the shape of the posterior distribution, and therefore lead to significantly different values for the total velocity of a star. We decide to adopt an \emph{exponentially decreasing prior}:
\begin{equation}
\label{eq:prior}
P(d) \propto d^2 \exp\Biggl(-\frac{d}{L}\Biggr),
\end{equation}
which has been shown to perform best for stars further out than $\sim 2$ kpc \citep{astraatmadjaII}, that is the expected distance of stars with a large relative error on parallax (see Appendix \ref{appendix:prior}). The value of the scale length parameter $L$ is fixed to $2600$ pc, and we refer the reader to the discussion in Appendix \ref{appendix:prior} for the reasons behind our choice of this particular value. By means of Bayes' theorem we can then express the posterior distribution on distances as:
\begin{equation}
\label{eq:post}
P(d|\varpi, \sigma_\varpi) \propto P(\varpi|d, \sigma_\varpi) P(d),
\end{equation}
where the likelihood probability $P(\varpi|d, \sigma_\varpi)$ is a Gaussian distribution centered on $1/d$:
\begin{equation}
\label{eq:likeli}
P(\varpi|d, \sigma_\varpi) \propto \exp\Biggl[-\frac{1}{2\sigma_\varpi^2}\Biggl(\varpi - \frac{1}{d}\Biggr)\Biggr].
\end{equation}
In our case, we decide to fully include the covariance matrix between the astrometric properties, following the approach introduced in \cite{marchetti+17}. In this case, for each star the likelihood probability is a three dimensional multivariate Gaussian distribution with mean vector:
\begin{equation}
\label{eq:mean_highf}
\mathbf{m} = [\mu_{\alpha*}, \mu_\delta, 1/d]
\end{equation}
and covariance matrix given by equation \eqref{eq:covmatr}. The prior distribution on distance is given by equation \eqref{eq:prior}, and we assume uniform priors on proper motions. We then draw proper motions and distances from the resulting posterior distribution using the affine invariant ensemble Markov chain Monte Carlo (MCMC) sampler \textsc{emcee} \citep{Goodman10, emcee}. We run each chain using $32$ walkers and $100$ steps, for a total of $3200$ random samples drawn from the posterior distribution. We initialize the walkers to random positions around the mean value of the proper motions and of the inverse of the mode of the posterior distribution in distance, equation \eqref{eq:post}, to achieve a fast convergence of the chain. We run $500$ burn-in steps to let the walkers explore the parameter space, and then we use the final positions as initial conditions for the proper MC chain. We then directly use this MC sampling to derive a distribution for the total velocity in the Galactic rest frame of each star, assuming the same parameters for the Sun presented in Section \ref{sec:lowf}. We check that the mean acceptance fraction (i.e. the fraction of steps accepted for each walker) is between $0.25$ and $0.5$ as a test for the convergence of each MC chain \citep{emcee}.

\section{The Total Velocity Distribution of Stars in \Gaia DR2}
\label{sec:v_distr}

\begin{figure}
	\centering
	\includegraphics[width=0.5\textwidth]{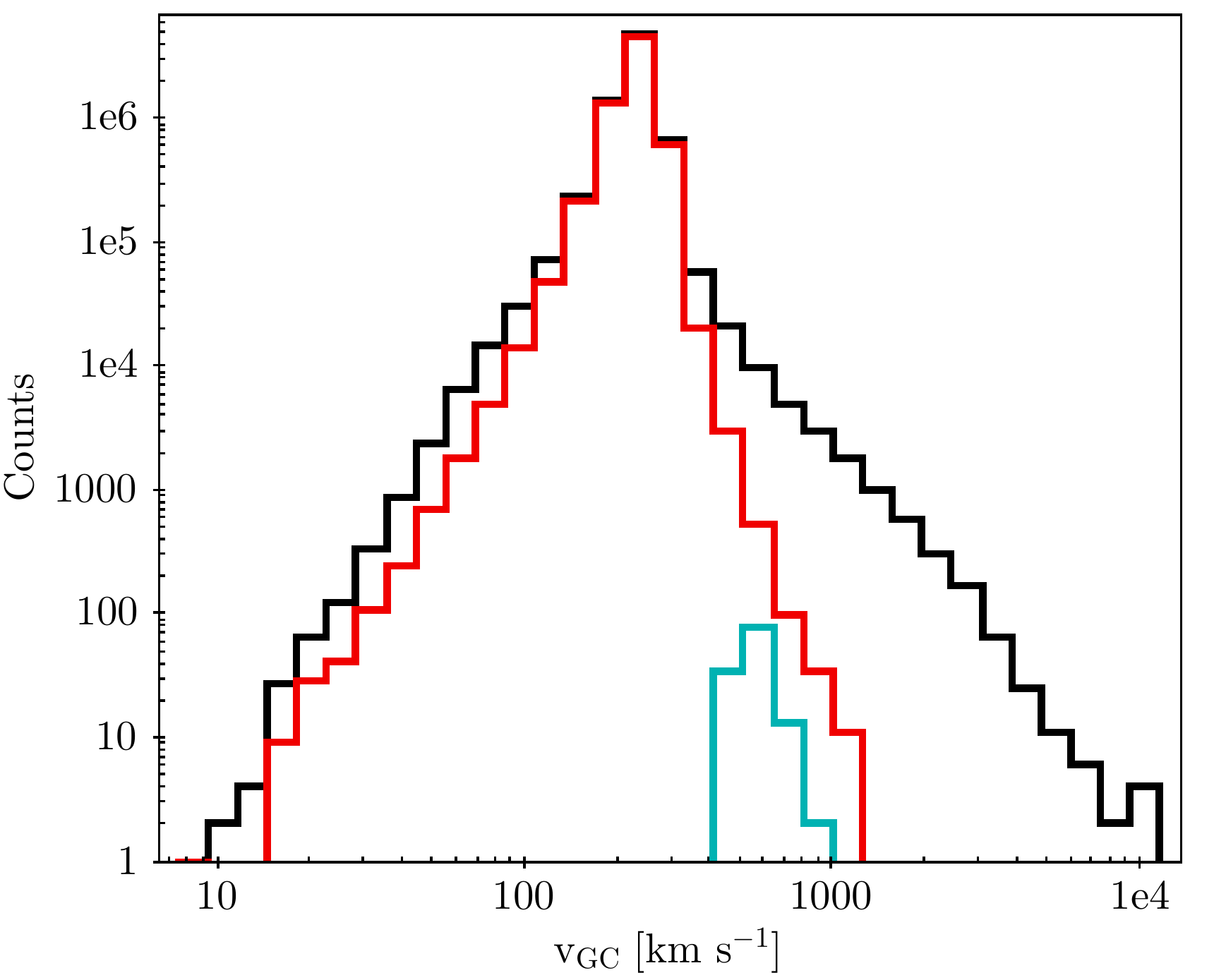}
	\caption{Histogram of median total velocities in the Galactic rest frame for all the $\sim 7$ million stars with three-dimensional velocity by \Gaia DR2 (black). The red line corresponds to those stars with a relative error on total velocity in the Galactic rest-frame below $30\%$, while the cyan line refers to our "clean" sample of high velocity stars (see discussion in Section \ref{sec:highV}).}
	\label{fig:vtot}
\end{figure}

\begin{figure*}
	\centering
	\includegraphics[width=\textwidth]{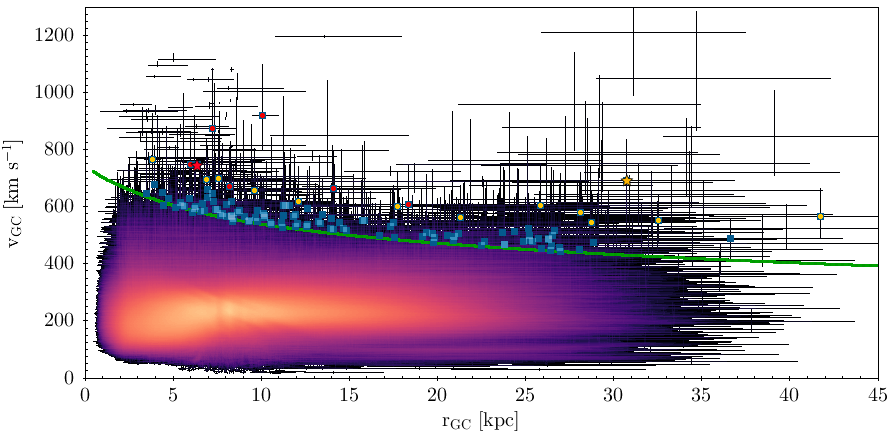}
	\caption{Total velocity in the Galactic rest-frame $v_\mathrm{GC}$ as a function of Galactocentric distance $r_\mathrm{GC}$ for all the $6884304$ stars in \Gaia DR2 with relative error on total velocity $< 0.3$. Colour is proportional to the logarithmic number density of stars. The green solid line is the median posterior escape speed from the adopted Galactic potential (Section \ref{sec:orbits}). We overplot in blue the "clean" high velocity star sample introduced in Section \ref{sec:highV}. Red and yellow points correspond, respectively, to the Galactic and extragalactic candidates discussed in Section \ref{sec:highVcand}. \Gaia DR2 5932173855446728064 (\Gaia DR2 1396963577886583296) is marked with a red (yellow) star.}
	\label{fig:rGC_vtot}
\end{figure*}

\begin{figure}
	\centering
	\includegraphics[width=0.5\textwidth]{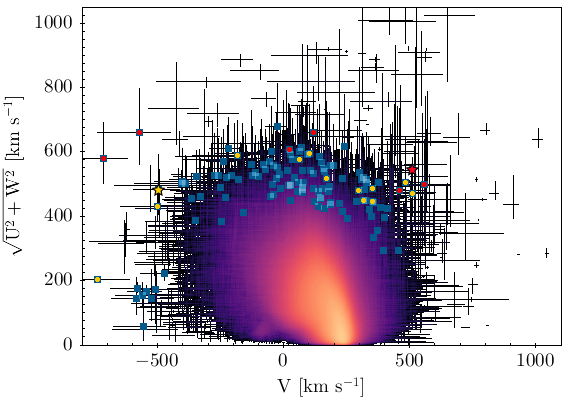}
	\caption{Toomre diagram for the same stars plotted in Fig. \ref{fig:rGC_vtot}.}
	\label{fig:Toomre}
\end{figure}

Using the approach discussed in Section \ref{sec:dist}, we publish a catalogue with distances and velocities in the Galactocentric frame for all the $7183262$ stars analyzed in this paper. This is publicly available at \url{http://home.strw.leidenuniv.nl/~marchetti/research.html}. A full description of the catalogue content can be found in Appendix \ref{appendix:catalogue}. 

In order to filter out the more uncertain candidates, for which it would be difficult to constrain the origin, we will now only discuss and plot results for stars with a relative error on total velocity $\sigma_{v_\mathrm{GC}} / v_\mathrm{GC} < 0.3$, where $\sigma_{v_\mathrm{GC}}$ is estimated summing in quadrature the lower and upper uncertainty on $v_\mathrm{GC}$. This cut results into a total of $6884304$ stars, $\sim 96\%$ of the original sample of stars. Figure \ref{fig:vtot} shows the total velocity distribution of the median Galactic rest frame total velocity $v_\mathrm{GC}$ for the original sample of $7183262$ stars (black line) and for the stars with a relative error on total velocity below $30 \%$ (red line). We can see how this cut filters out most of the stars with extremely high velocities, which are likely to be outliers with relatively more uncertain measurements by \Gaia. Nevertheless we note the presence of a high velocity tail extending up to and above $\sim 1000$ \kms surviving the cut. We will now focus only on stars with $\sigma_{v_\mathrm{GC}} / v_\mathrm{GC} < 0.3$.

To highlight visually possibly unbound objects, we plot in Figure \ref{fig:rGC_vtot} the total velocity for all stars as a function of the Galactocentric distance $r_\mathrm{GC}$, and we overplot the median escape speed from the Galaxy with a green solid line, computed using the Galactic potential model introduced in Section \ref{sec:orbits}. Datapoints correspond to the medians of the distributions, with lower and upper uncertainties derived, respectively, from the $16$th and $84$th percentiles. Most of the stars are located in the solar neighborhood, and have typical velocities of the order of the LSR velocity. We find $510$ stars to have probabilities greater than $50 \%$ of being unbound from the Galaxy (but note the large errorbars). In particular, $212$ ($103$) stars are more than $1$-$\sigma$ ($3$-$\sigma$) away from the Galactic escape speed.

Figure \ref{fig:Toomre} shows the Toomre diagram for all the $\sim 7$ million stars, a plot that is useful to distinguish stellar populations based on their kinematics. On the $x$-axis we plot the component $V$ of the Galactocentric Cartesian velocity, and on the $y$-axis the component orthogonal to it, $\sqrt{U^2 + W^2}$. Not surprisingly, most of the stars behave kinematically as disk stars on rotation-supported orbits, with $V$ values around the Sun's orbital velocity \citep[see][]{gaiadisk}. A sub-dominant, more diffuse, population of stars with halo-like kinematics is also present, centered around $V = 0$ and with a larger spread in total velocity. 

\section{High Velocity Stars in \Gaia DR2}
\label{sec:highV}

\begin{figure}
	\centering
	\includegraphics[width=0.5\textwidth]{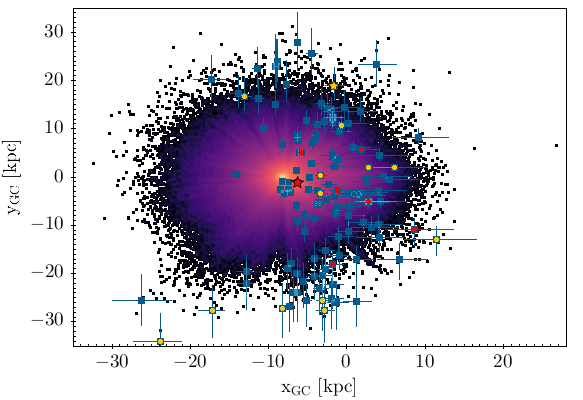}
	\caption{Distribution of the $\sim 7$ million stars on the Galactic plane. The Sun is located at $(x_\mathrm{GC}, y_\mathrm{GC}) = (-8.2, 0)$ kpc. Colours are the same as in Fig. \ref{fig:rGC_vtot}.}
	\label{fig:xGC_yGC}
\end{figure}

\begin{figure}
	\centering
	\includegraphics[width=0.5\textwidth]{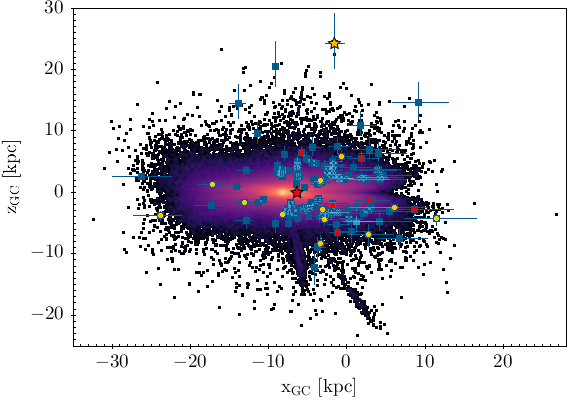}
	\caption{Same as Fig. \ref{fig:xGC_yGC}, but showing the distribution of the stars in the $(x_\mathrm{GC}, z_\mathrm{GC})$ plane. The Sun is located at $(x_\mathrm{GC}, z_\mathrm{GC}) = (-8200, 25)$ pc. Colors are the same as in Fig. \ref{fig:rGC_vtot}.}
	\label{fig:xGC_zGC}
\end{figure}

\begin{figure}
	\centering
	\includegraphics[width=0.5\textwidth]{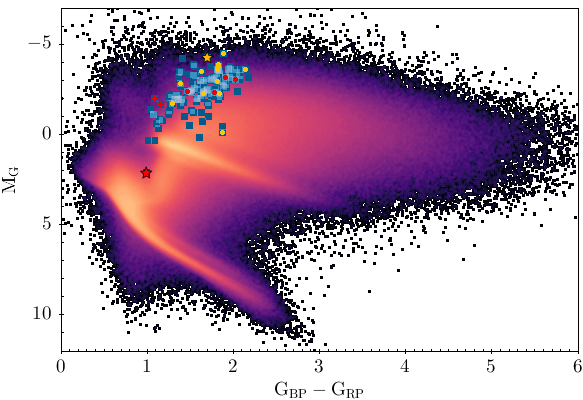}
	\caption{HR diagram for all the $\sim 7$ million stars in \Gaia DR2 with a radial velocity measurement. Colours are the same as in Fig. \ref{fig:rGC_vtot}.}
	\label{fig:HR}
\end{figure}

We now focus our interest towards high velocity stars, which we define as stars with a probability $P_\mathrm{ub} > 0.5$. Since we are interested in kinematic outliers, we have to pay particular attention not to be contaminated by data processing artifacts and/or spurious measurements. We therefore choose to adopt the following conservative cuts on the columns of the \Gaia DR2 \textsc{gaia\_source} catalogue (in addition to the selection $\sigma_{v_\mathrm{GC}} / v_\mathrm{GC} < 0.3$ introduced in Section \ref{sec:v_distr}):

\begin{enumerate}

\item \textsc{astrometric\_gof\_al} $ < 3$;
\item \textsc{astrometric\_excess\_noise\_sig} $\le 2$;
\item $-0.23 \le$ \textsc{mean\_varpi\_factor\_al} $\le 0.32$;
\item \textsc{visibility\_periods\_used} $>8$;
\item \textsc{rv\_nb\_transits} $>5$.

\end{enumerate}

The first cut ensures that statistic astrometric model resulted in a good fit to the data, while the second cut selects only astrometrically well-behaved sources \citep[refer to][for a detailed explanation of the excess noise and its significance]{lindegren+12}. The third and the fourth cuts are useful to exclude stars with parallaxes more vulnerable to errors. Finally, the final selection ensures that each source was observed a reasonable number of times ($5$) by \Gaia to determine its radial velocity. Further details on the parameters used to filter out possible contaminants and the reasons behind the adopted threshold values can be found in the \Gaia data model\footnote{\url{https://gea.esac.esa.int/archive/documentation/GDR2/Gaia_archive/chap_datamodel/}}. Applying these cuts and with the further constrain on the unbound probability $P_\mathrm{ub} > 0.5$, we are left with a clean final sample of $125$ high velocity stars. We also verify that the quality cuts C.1 and C.2 introduced in Appendix C of \cite{lindegren+18}, designed to select astrometrically clean subsets of objects, are already verified by our sample of high velocity stars. In addition, selection N in Appendix C of \cite{lindegren+18} does not select any of our candidates. Looking at Fig. \ref{fig:rGC_vtot}, where this clean sample of 125 stars is highlighted with blue squares, we can see how these cuts filter out most of the stars with exceptionally high velocities, which are therefore likely to be instrumental artifacts. This is also evident in Fig. \ref{fig:vtot}, where the Galactic rest-frame total velocity distribution of the 125 high velocity stars is shown with a cyan line.

We present distances, total velocities, and probability of being unbound for all the $105$ stars wih $0.5 < P_\mathrm{ub} \leq 0.8$ in Appendidx \ref{appendix:candidates_0.5}, Table \ref{tab:candidates_0.5}. Stars with $P_\mathrm{ub} > 0.8$ are presented and discussed in detail in Section \ref{sec:highVcand}.

The spatial distribution of these $125$ high velocity stars in our Galaxy is shown in Fig. \ref{fig:xGC_yGC}, where we overplot the position on the Galactic plane of this subset of stars with blue markers above the underlying distribution of the $\sim 7$ million stars used in this paper. We can see how the majority of high velocity stars lies in the inner region of the Galaxy, with typical distances $\lesssim 15$ kpc from the GC. Most of these stars are on the faint end of the magnitude distribution because of extinction due to dust in the direction of the GC, and thus they have large relative errors on parallax. This in turn translates into larger uncertainties on total velocity, which may cause the stars to be included into our high velocity cut. Another small overdensity corresponds to the Sun's position, correlating with the underlying distribution of all the stars. In Fig. \ref{fig:xGC_zGC}, we plot the same but in the $(x_\mathrm{GC}, z_\mathrm{GC})$ plane. Most of our high velocity stars lie away from the stellar disk.

Fig. \ref{fig:HR} shows the Hertzsprung-Russell (HR) diagram for all the sources with a radial velocity measurement, with the high velocity star sample overplotted in blue. On the $x$-axis we plot the color index in the \Gaia Blue Pass (BP) and Red Pass (RP) bands $G_\mathrm{BP} - G_\mathrm{RP}$, while on the $y$-axis we plot the absolute magnitude in the \Gaia $G$ band $M_G$, computed assuming the median of the posterior distance distribution. Note that we did not consider extinction to construct the HR diagram, because of the caveats with using the line-of-sight extinction in the $G$ band $A_G$ for individual sources \citep{andrae+18}. We can see that the great majority of our stars are giants stars. This is consistent with recent findings of \cite{hattori+18b,hawkins+18}, which confirm some of these candidates as being old ($ > 1$ Gyr), metal-poor giants ($2 \leq$ [Fe/H] $\leq 1$).

\subsection{Orbital Integration}
\label{sec:orbits}

\begin{figure}
	\centering
 	\includegraphics[width=0.5\textwidth]{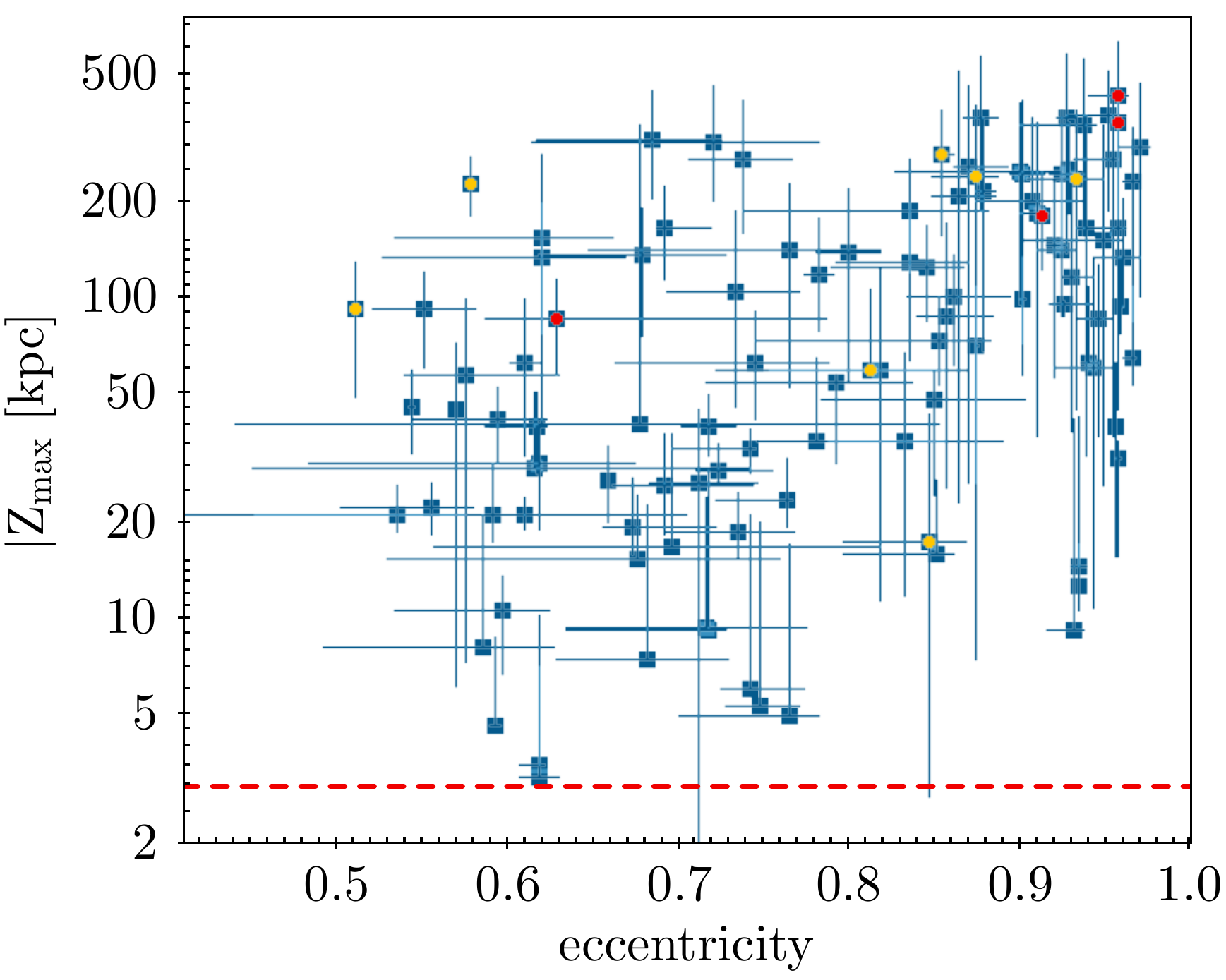}
	\caption{Absolute value of the maximum height above the Galactic plane $|Z_\mathrm{max}|$ as a function of eccentricity for the high velocity sample of stars. The yellow horizontal dashed line corresponds to $Z_\mathrm{max} = 3$ kpc, the edge of the thick disk \citep{carollo+10}. Colours are the same as in Fig. \ref{fig:rGC_vtot}.}
	\label{fig:ecc_zmax}
\end{figure}

In order to get hints on the ejection location of our sample of high velocity stars, we perform numerical orbit integration of their trajectories back in time using the python package \textsc{Gala} \citep{gala}. For each star we use $1000$ random samples from the proper motions, distance, and radial velocity MC sampling discussed in Section \ref{sec:dist}. We integrate each orbit back in time for a total time of $1$ Gyr, with a fixed time-step of $0.1$ Myr, using the \textsc{gala} potential \emph{MilkyWayPotential}. This is a four components Galactic potential model consisting of a Hernquist bulge and nucleus \citep{hernquist90}:
\begin{equation}
\label{eq:HernquistBulge}
\phi_b(r_\mathrm{GC}) = -\frac{G M_i}{r_\mathrm{GC} + r_i},
\end{equation}
where $i = b, n$ for the bulge and the nucleus, respectively, a Miyamoto-Nagai disk \citep{M&N75}:
\begin{equation}
\label{eq:MNdisk}
\phi_d(R_\mathrm{GC}, z_\mathrm{GC}) = -\frac{G M_d}{\sqrt{R_\mathrm{GC}^2 + \Bigl(a_d + \sqrt{z_\mathrm{GC}^2 + b_d^2}\Bigr)^2}},
\end{equation}
and a Navarro-Frenk-White halo \citep{nfw96}:
\begin{equation}
\label{eq:NFW}
\phi_h(r_\mathrm{GC}) = -\frac{G M_h}{r_\mathrm{GC}} \ln \Bigl(1 + \frac{r_\mathrm{GC}}{r_s}\Bigr).
\end{equation}
The parameters are chosen to fit the enclosed mass profile of the Milky Way \citep{bovy15}, and are summarized in Table \ref{tab:potential}. We then derive the pericenter distance and, for bound MC realizations, the apocenter distance and the eccentricity of the orbit. We also record the energy and the angular momentum of each MC orbit. We check for energy conservation as a test of the accuracy of the numerical integration.

\begin{table}
\centering
\caption{Parameters for the \textsc{Gala} potential \emph{MilkyWayPotential}.}
\label{tab:potential}
\begin{tabular}{l|l}
	\hline
    Component & Parameters \\
    \hline
    Bulge & $M_b = 5.00 \cdot 10^9$ M$_\odot$ \\
     & $r_b = 1.00$ kpc \\
    Nucleus & $M_n = 1.71 \cdot 10^9$ M$_\odot$ \\
     & $r_n = 0.07$ kpc \\
    Disk & $M_d = 6.80 \cdot 10^{10}$ M$_\odot$ \\
     & $a_d = 3.00$ kpc \\
     & $b_d = 0.28$ kpc \\
    Halo & $M_h = 5.40 \cdot 10^{11}$ M$_\odot$ \\
     & $r_s = 15.62$ kpc \\
    \hline
\end{tabular}
\end{table}

\begin{figure}
	\centering
	\includegraphics[width=0.5\textwidth]{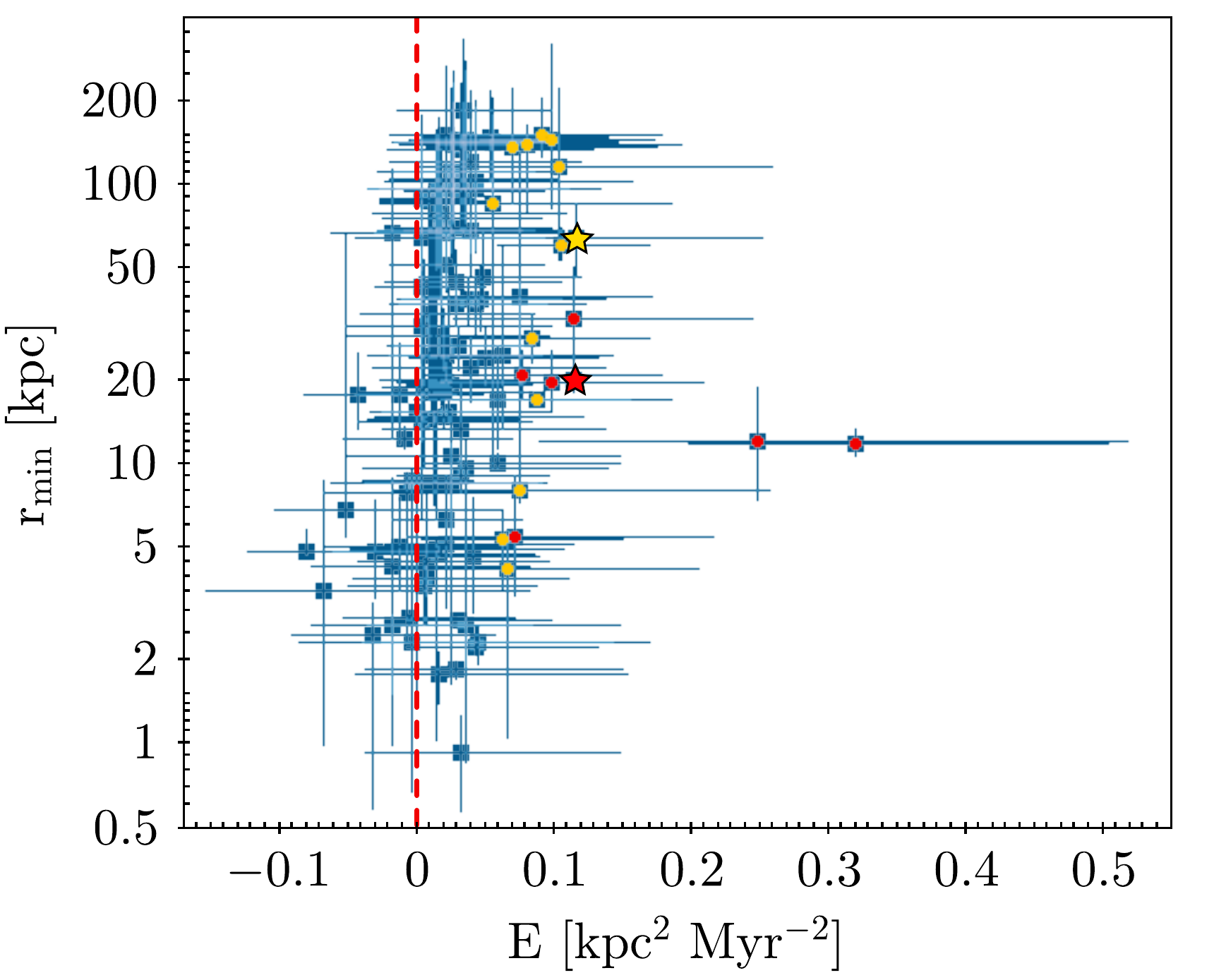}
	\caption{Minimum crossing radius $r_\mathrm{min}$ versus energy $E$ for the $125$ high velocity stars. The vertical dashed line separates unbound ($E>0$) from bound ($E<0$) orbits. Colors are the same as in Fig. \ref{fig:rGC_vtot}.}
	\label{fig:rmin_E}
\end{figure}

In Fig. \ref{fig:ecc_zmax}, we plot the maximum height above the Galactic disk $Z_\mathrm{max}$ as a function of the eccentricity of the orbit for our sample of high velocity stars. This plot is useful to identify similar stars based on their orbits \citep[e.g.][]{boeche+13, hawkins+15}. The dashed red line at $Z_\mathrm{max} = 3$ kpc denotes the typical scale height of the thick disk \citep{carollo+10}. Not surprisingly, high velocity stars are on highly eccentric orbits, with a mean eccentricity of the sample $\sim 0.8$. Most of these stars span a large range of $Z_\mathrm{max}$, with values up to hundreds of kpc, reflecting the large amplitude of the vertical oscillations. 

In our search for HVSs, we keep track of each disk crossing (Cartesian Galactocentric coordinate $z_\mathrm{GC} = 0$) in the orbital traceback of our high velocity star sample. For each MC realization, we then define the crossing radius $r_c$ as:
\begin{equation}
\label{eq:rcross}
r_c = \sqrt{x_c^2 + y_c^2},
\end{equation}
where $x_c$ and $y_c$ are the Galactocentric coordinates of the orbit $(x_\mathrm{GC}, y_\mathrm{GC})$ at the instant when $z_\mathrm{GC} = 0$. In the case of multiple disk crossings during the orbital trace-back, we define $r_\mathrm{min}$ as the minimum crossing radius attained in that particular MC realization of the star's orbit. This approach allows us to check for the consistency of the GC origin hypothesis for our sample of high velocity stars. We also record the ejection velocity $v_\mathrm{ej}$: the velocity of the star at the minimum crossing radius, and the flight time $t_\mathrm{f}$: the time needed to travel from the observed position to the disk crossing happening closest to the GC.

In Fig. \ref{fig:rmin_E}, we plot $r_\mathrm{min}$ as a function of the orbital energy $E$. The red dashed line coincides with the separation region between bound and unbound orbits. The majority of candidates are traveling on unbound orbits ($E > 0$), and we can see a few stars with remarkably high values of the energy: $25$ stars are unbound at more than $1$ sigma significance, and $1$ star (\Gaia DR2 5932173855446728064) is unbound at more than $3$ sigma significance.

\section{Unbound Stars: Hypervelocity and Hyper Runaway Star Candidates}
\label{sec:highVcand}

\begin{figure*}
	\centering
	\includegraphics[width=0.75\textwidth]{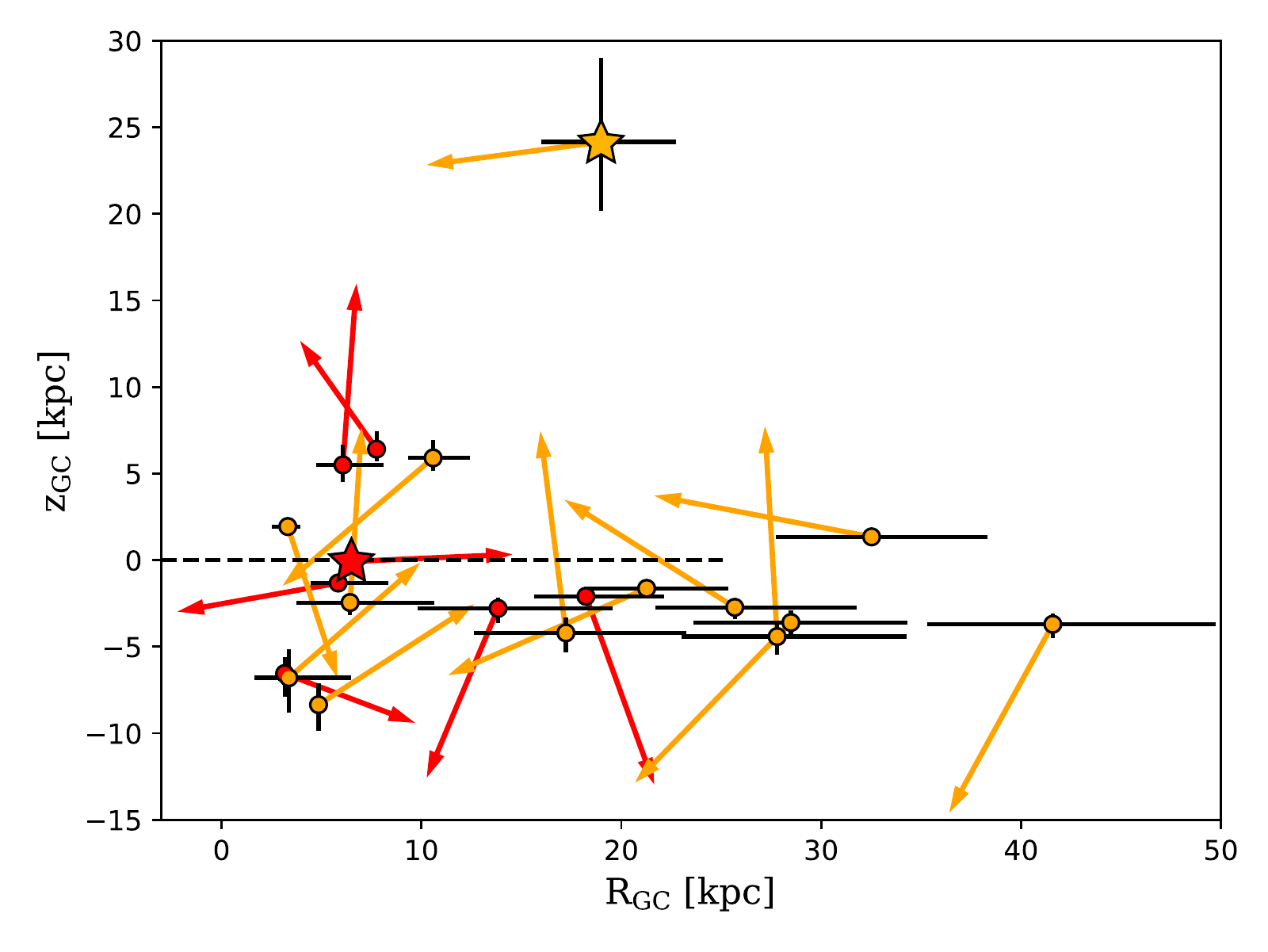}
	\caption{Position of the $20$ high velocity stars with $P_\mathrm{ub} > 80 \%$ in Galactocentric cylindrical coordinates $(R_\mathrm{GC}, z_\mathrm{GC})$. Arrows point to the direction of the velocity vector of the stars in this coordinate system, and the arrow's length is proportional to the total velocity of the star in the Galactic rest-frame. Red (yellow) points and arrows mark the 7 (13) Galactic (extragalactic) candidates with $P_\mathrm{MW} > 0.5$ ($P_\mathrm{MW} < 0.5$). \Gaia DR2 5932173855446728064 (\Gaia DR2 1396963577886583296) is marked with a red (yellow) star. The Sun is located at $(R_\mathrm{GC}, z_\mathrm{GC}) = (8200, 25)$ pc. The horizontal dashed line denotes the position of the Galactic plane, and extends up to the edge of the stellar disk, which we take to be at 25 kpc \citep{xu+15}.}
	\label{fig:RGC_zGC}
\end{figure*}

\begin{figure}
	\centering
	\includegraphics[width=0.5\textwidth]{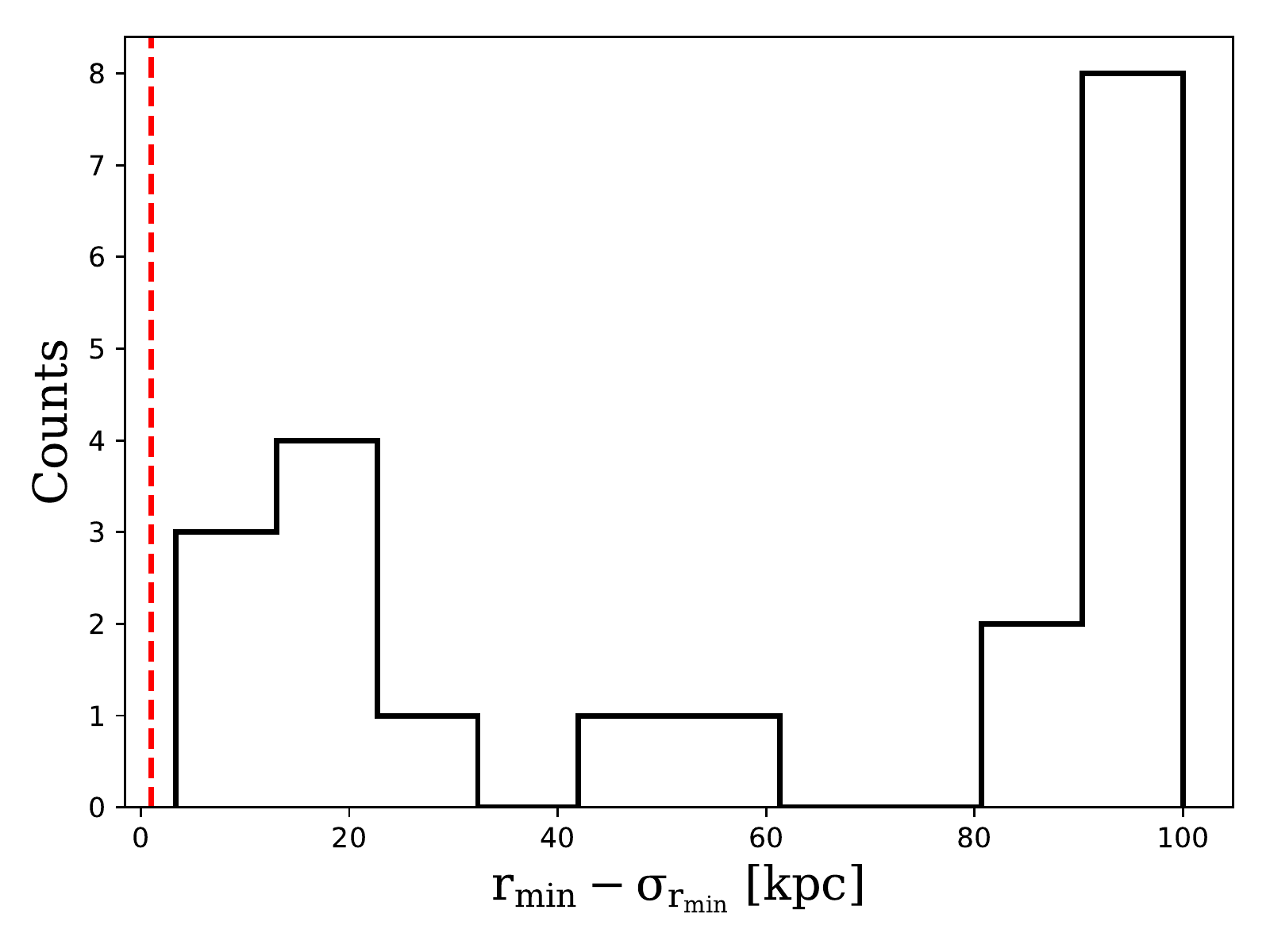}
	\caption{Histogram of the median minimum crossing radius $r_\mathrm{min}$ minus the correspondent lower uncertainty $\sigma_{r_\mathrm{min, l}}$ for the sample of $20$ high velocity stars with $P_\mathrm{ub} > 0.8$. The vertical dashed line corresponds to $(r_\mathrm{min} - \sigma_{r_\mathrm{min, l}} ) = 1$ kpc, our boundary condition for not rejecting the GC origin hypothesis for the HVS candidates (see discussion in Section \ref{sec:highVcand}). $(r_\mathrm{min} - \sigma_{r_\mathrm{min, l}} ) > 1$ kpc for all the $20$ stars, therefore there are no HVS candidates.}
	\label{fig:rmin}
\end{figure}

We now focus our search on possible unbound stars, defined as the subsample of clean high velocity stars with $P_\mathrm{ub} > 80 \%$. This amounts to a total of $20$ objects. Observed properties from \Gaia DR2, distances, and total velocities for these stars are summarized in Table \ref{tab:candidates}. Fig. \ref{fig:RGC_zGC} shows the position in Galactocentric cylindrical coordinates of these high velocity star candidates. The length of the arrows is proportional to the total velocity of each star in the Galactic rest frame. We note that for most of our candidates ($18$ out of $20$ stars) the parallax uncertainty is smaller than the quoted parallax zeropoint of $-0.029$ mas, as estimated by \Gaia's observations of quasars \citep{lindegren+18}. We discuss the impact of considering this negative offset in the analysis of our stars in Appendix \ref{appendix:offset}. We further discuss the impact of systematic errors for our sample of $20$ unbound candidates in Appendix \ref{appendix:inflated_erros}.

If a star on an unbound orbit was ejected either from the stellar disk (HRS) or from the GC (HVS), then its distribution of minimum crossing radii $r_\mathrm{min}$ should fall within the edge of the Milky Way disk. To maximize the probability of a disk crossing during the orbital traceback, we integrate the orbits of these stars for a maximum time of $5$ Gyr. We then define the probability $P_\mathrm{MW}$ for a star to come from the Milky Way as the fraction of MC realizations resulting in a minimum crossing radius within the edge of the stellar disk: $r_\mathrm{min} < r_\mathrm{disk}$, where $r_\mathrm{disk} = 25$ kpc \citep{xu+15}. This probability is useful to flag candidates of possible extragalactic origin, which we define as those stars with $P_\mathrm{MW} < 0.5$. This subset of $13$ stars, if their high velocity is confirmed, could either originate as RS/HRS/HVS from the LMC \citep{boubert+16,boubert+17a,erkal+18}, or could be the result of the tidal disruption of a dwarf galaxy interacting with the Milky Way \citep{abadi+09}. Stars with a Galactic and extragalactic origin are marked in Fig. \ref{fig:RGC_zGC} with red and yellow points, respectively. Stars with a Galactic origin have trajectories pointing away from the stellar disk. On the other hand, extragalactic stars are pointing either towards the disk, or are consistent with coming from regions of no current active star formation (i.e. the outer halo).

\subsection{Galactic Stars}
\label{sec:galactic}

$7$ of the $20$ possible unbound stars have $P_\mathrm{MW} > 0.5$, and therefore are consistent with being ejected from the stellar disk of the Milky Way. These stars, given their extremely high velocities, could be either HVS or HRS candidates. 

We then classify a star as a HVS (HRS) candidate if we cannot (can) exclude the hypothesis of GC origin, which we define by the condition $r_\mathrm{min} - \sigma_{r_\mathrm{min, l}} < 1$ kpc ($r_\mathrm{min}  - \sigma_{r_\mathrm{min, l}} > 1$ kpc), where $r_\mathrm{min}$ denotes the median of the distribution, and $\sigma_{r_\mathrm{min, l}}$ is the lower uncertainty on the minimum crossing radius. In this way we are testing whether, within its errorbars, a star is consistent with coming from the central region of the Galaxy. Figure \ref{fig:rmin} shows the histogram of the median minimum disk crossing $r_\mathrm{min}$ minus the lower uncertainty $\sigma_{r_\mathrm{min, l}}$ for all the $20$ stars with $P_\mathrm{ub} > 0.8$. A vertical red dashed line corresponds to the value $1$ kpc, which we use to define HVS candidates.

We find that all of these $7$ stars have orbits that, when integrated back in time, are not consistent with coming from the GC. Therefore, according to our classification criterion, there are no stars classified as HVS candidates. The absence of HVS candidates in the subset of \Gaia DR2 with radial velocities was anticipated by predictions by \cite{marchetti+18}, analyzing the \textsc{Hills} mock catalogue of HVSs. This is due to the fact that the expected number density of HVSs generated via the Hills' mechanism is expected to increase linearly with increasing galactocentric distance \citep{brown15}, and the majority of HVSs in the Milky Way are too faint to have a radial velocity measurement from \Gaia DR2. We cannot exclude the presence of \emph{bound} HVSs in the subset of $\sim 7$ million stars considered in this work, but their identification is not trivial because of their complex orbits and lower velocities. About 20 BHVSs are expected to have radial velocities from \Gaia DR2 \citep{marchetti+18}, but their identification is beyond the scope of this manuscript.

All the $7$ Galactic stars are therefore HRS candidates (red circles in Fig. \ref{fig:rGC_vtot} and following plots). One particular HRS candidate that is worth mentioning is \Gaia DR2 5932173855446728064 (marked with a red star in Fig. \ref{fig:rGC_vtot} and following). This star has an exceptionally well constrained total velocity\footnote{Because of the small uncertainties, we repeat the total velocity determination for \Gaia DR2 5932173855446728064 sampling within the uncertainties of the Sun position and motion (see discussion in Section \ref{sec:lowf}). The result is $v_\mathrm{GC} = (747 \pm 7)$ \kms, in agreement with the previous estimate.}, $v_\mathrm{GC} = 747^{+2}_{-3}$ \kms, which results in a probability of being unbound $\approx 1$. This star most likely was ejected in the thin disc of the Milky Way.

We note that $5$ of the $7$ HRS candidates with a Galactic origin have $P_\mathrm{ub} > 90\%$. Such exceptionally high velocities are thought to be very uncommon in our Galaxy for HRSs, which are predicted to be much rarer than HVSs \citep{brown15}. This is correct in the context of the Milky Way as a whole. In this study we only focus on bright sources ($G_\mathrm{RVS} < 12$), therefore we maximize the probability of observing stars ejected from the stellar disk. The HVS population is instead expected to be much fainter than this magnitude cut \citep{marchetti+18}. Since estimates on the expected HRS population in \Gaia are currently missing, at the moment it is not clear whether this tension is real, and/or if other ejection mechanisms are needed \cite[e.g.][]{irrgang+18}.

\subsection{Extragalactic Stars}
\label{sec:extragalactic}

$13$ of the $20$ $P_\mathrm{ub} > 80$\% stars have probabilities $< 50 \%$ of intersecting the Milky Way stellar disk when traced back in time, therefore an extragalactic origin is preferred. A possible ejection location could be the LMC, or otherwise spatial correlations with the density of surrounding stars could help identifying them as the high velocity tail of a stellar stream produced by the effect of the gravitational field of the Milky Way on a dwarf satellite galaxy \citep{abadi+09}. 

The extragalactic star with a highest probability of being unbound from our Galaxy is Gaia DR2 1396963577886583296, with a total velocity $\sim 700$ \kms, resulting in a probability $P_\mathrm{ub} = 0.98$. We mark this source with a yellow star in Fig. \ref{fig:rGC_vtot} and following. This star is at $\sim 30$ kpc from the GC, with an elevation of $\sim 25$ kpc above the Galactic plane.

\section{Conclusions} 
\label{sec:discussion}

We derived distance and total velocities for all the $7183262$ stars with a full phase space measurement in the \Gaia DR2 catalogue, in order to find unbound objects and velocity outliers. We defined our sample of high velocity stars as those stars with an estimated probability of being unbound from the Milky Way $P_\mathrm{ub} > 50\%$, resulting in a total of $125$ stars with reliable astrometric parameters and radial velocities. We traced back the high velocity stars in the Galactic potential to derive orbital parameters. Out of these $125$ stars, we found the following.

\begin{enumerate}

\item $20$ stars have predicted probabilities $P_\mathrm{ub} > 80\%$. The observed and derived kinematic properties of these stars are summarized in Table \ref{tab:candidates}, and are discussed in Section \ref{sec:highVcand}.

\item None of these 20 stars is consistent with coming from the inner $1$ kpc, so there are no HVS candidates. This is consistent with estimates presented in \cite{marchetti+18}.

\item $7$ out of the $20$ stars with $P_\mathrm{ub} > 0.8$, when traced back in time in the Galactic potential, originate from the stellar disk of the Milky Way. These stars are HRS candidates.

\item $13$ out of the $20$ unbound candidates have probabilities $< 50\%$ to originate from the stellar disk of the Galaxy. This surprising and unexpected population of stars could be either produced as RSs/HRSs/HVSs from the LMC, thanks to its high orbital velocity around the Milky Way, or could be members of dwarf galaxies tidally disrupted by the gravitational interaction with the Galaxy. Further analyses are required in order to identify their origin.

\end{enumerate}

Another possibility that we cannot rule out is that a subset of these $20$ stars is actually gravitationally \emph{bound} to the Milky Way. Recent high-resolution spectroscopic followups showed that some of these stars are actually indistinguishable from halo stars from a chemical point of view \citep{hawkins+18}, therefore if they are actually bound, this would in turn imply a more massive Milky Way \citep{hattori+18b,monari+18}, a possiblity that cannot be ruled out \citep[e.g.][]{wang+15}. Otherwise, a confirmation of the global parallax zeropoint measured with quasars could lower down their total velocities, resulting in the same effectAs discussed in Appendix \ref{appendix:offset}, including this parallax offset results in 14 (4) stars with an updated $P_\mathrm{ub} > 50\%$ ($P_\mathrm{ub} > 80\%$). The choice of not considering the parallax zero point in the main text is therefore a conservative choice, which ensures us that all the high velocity stars in the subset of \Gaia DR2 with radial velocities are actually included in this work. In Appendix \ref{appendix:inflated_erros} we show how including systematic errors in parallax can significantly lower the distances and total velocities for our candidates, but we want to stress that the adopted parameters might be too pessimistic for the stars considered in this paper \citep{Lindegren+18p}. Follow-up observations with ground based facilities and/or future data releases of the \Gaia satellite will help us confirming or rejecting their interpretation as kinematic outliers.

This paper is just a first proof of the exciting discoveries that can be made mining the \Gaia DR2 catalogue. We only limited our search to the $\sim 7$ million stars with a full phase space information, a small catalogue compared to the full $1.3$ billion sources with proper motions and parallaxes. Synergies with existing  and upcoming ground-based spectroscopic surveys will be essential to obtain radial velocities and stellar spectra for subsets of these stars \citep[e.g.][]{dalton16, dejong+16, kunder+16, galah}. For what concerns HVSs, \cite{marchetti+18} shows how the majority of HVSs expected to be found in the \Gaia catalogue are actually fainter than the limiting magnitude for radial velocities in DR2. We therefore did not expect to discover the bulk of the HVS population with the method outlined in this paper, but other data mining techniques need to be implemented in order to identify them among the dominant background of bound, low velocity stars \citep[see for example][]{marchetti+17}. We also show how particular attention needs to be paid to efficiently filter out contaminants and instrumental artifacts, which might mimic high velocity stars at a first inspection.

\section*{Acknowledgements}

We thank the anonymous referee for his/her comments, which greatly improved the quality of this manuscript. We also thank E. Zari and the \Gaia group meeting at Leiden Observatory for useful comments, suggestions and discussions during the preparation of this paper. TM and EMR acknowledge support from NWO TOP grant Module 2, project number 614.001.401. This project was developed in part at the 2017 Heidelberg \Gaia Sprint, hosted by the Max-Planck-Institut f\"{u}r Astronomie, Heidelberg. This work has made use of data from the European Space Agency (ESA) mission \Gaia (\url{https://www.cosmos.esa.int/gaia}), processed by the \Gaia Data Processing and Analysis Consortium (DPAC, \url{https://www.cosmos.esa.int/web/gaia/dpac/consortium}). Funding for the DPAC has been provided by national institutions, in particular
the institutions participating in the \Gaia Multilateral Agreement. This research made use of \textsc{Astropy}, a community-developed core \textsc{Python} package for Astronomy \citep{astropy}. All figures in the paper were produced using \textsc{matplotlib} \citep{matplotlib} and \textsc{Topcat} \citep{topcat}. This work would not have been possible without the countless hours put in by members of the open-source community all around the world.

\begin{landscape}
\begin{table}
\centering
\caption{Observed and derived properties for the $20$ "clean" high velocity star candidates with a probability $> 80\%$ of being unbound from the Galaxy. Stars are sorted by decreasing $P_\mathrm{ub}$.}
\label{tab:candidates}

\begin{threeparttable}

\begin{tabular}{lccccccccccc}

\hline
\Gaia DR2 ID & (RA, Dec.) & $\varpi$ & $\mu_{\alpha*}$ & $\mu_\delta$ & $v_\mathrm{rad}$ & $G$ & $d$ & $r_\mathrm{GC}$ & $v_\mathrm{GC}$ & $P_\mathrm{MW}$ & $P_\mathrm{ub}$ \\
 & ($^\circ$) & (mas) & (mas yr$^{-1}$) & (mas yr$^{-1}$) & (km s$^{-1}$) & (mag) & (pc) & (pc) & (\kms) &  &  \\
\hline

\textbf{Galactic} \\[0.15cm]

5932173855446728064 & ($244.1181$, $-54.44045$) & $0.454 \pm 0.029$ & $-2.676 \pm 0.043$ & $-4.991 \pm 0.034$ & $-614.286 \pm 2.492$ & $13.81$ & $2197^{+162}_{-120}$ & $6397^{+92}_{-123}$ & $747^{+2}_{-3}$ & $1.00$ & $1.00$ \\[0.15cm] 
1383279090527227264 & ($240.33735$, $41.16677$) & $0.118 \pm 0.016$ & $-25.759 \pm 0.025$ & $-9.745 \pm 0.04$ & $-180.902 \pm 2.421$ & $13.01$ & $8491^{+1376}_{-951}$ & $10064^{+908}_{-561}$ & $921^{+179}_{-124}$ & $1.00$ & $1.00$ \\[0.15cm]
6456587609813249536 & ($317.36089$, $-57.9124$) & $0.099 \pm 0.019$ & $13.002 \pm 0.029$ & $-18.263 \pm 0.03$ & $-15.851 \pm 2.833$ & $13.01$ & $10021^{+2023}_{-1480}$ & $7222^{+1350}_{-761}$ & $875^{+212}_{-155}$ & $0.98$ & $0.99$ \\[0.15cm] 
5935868592404029184 & ($253.90291$, $-53.29868$) & $0.074 \pm 0.021$ & $5.47 \pm 0.032$ & $6.358 \pm 0.026$ & $308.412 \pm 1.212$ & $13.08$ & $12150^{+2919}_{-1909}$ & $5985^{+2516}_{-1380}$ & $747^{+110}_{-73}$ & $0.83$ & $0.98$ \\[0.15cm] 
5831614858352694400 & ($247.45238$, $-59.96738$) & $-0.008 \pm 0.025$ & $4.405 \pm 0.032$ & $1.532 \pm 0.03$ & $258.295 \pm 1.245$ & $13.37$ & $20196^{+6006}_{-4394}$ & $14113^{+5781}_{-4061}$ & $664^{+130}_{-93}$ & $0.94$ & $0.92$ \\[0.15cm]
5239334504523094784 & ($158.89457$, $-65.46548$) & $0.038 \pm 0.013$ & $-6.77 \pm 0.025$ & $2.544 \pm 0.022$ & $22.464 \pm 1.891$ & $13.39$ & $19353^{+4247}_{-2940}$ & $18351^{+3923}_{-2617}$ & $609^{+140}_{-94}$ & $0.77$ & $0.88$ \\[0.15cm] 
4395399303719163904 & ($258.75009$, $8.73145$) & $0.073 \pm 0.019$ & $-9.911 \pm 0.029$ & $4.848 \pm 0.029$ & $24.364 \pm 1.484$ & $13.19$ & $12848^{+2766}_{-2262}$ & $8194^{+2309}_{-1620}$ & $671^{+136}_{-106}$ & $1.00$ & $0.84$ \\[0.15cm]

\textbf{Extragalactic} \\[0.15cm]

1396963577886583296 & ($237.73164$, $44.4357$) & $-0.017 \pm 0.014$ & $-1.649 \pm 0.023$ & $-4.966 \pm 0.029$ & $-412.464 \pm 1.002$ & $13.24$ & $31374^{+6332}_{-5185}$ & $30720^{+6150}_{-4970}$ & $693^{+145}_{-113}$ & $0.00$ & $0.98$ \\[0.15cm]
5593107043671135744 & ($113.26944$, $-31.3792$) & $-0.1 \pm 0.017$ & $-1.582 \pm 0.03$ & $2.113 \pm 0.028$ & $104.437 \pm 1.511$ & $13.39$ & $37681^{+8295}_{-6444}$ & $41753^{+8183}_{-6322}$ & $567^{+100}_{-76}$ & $0.00$ & $0.97$ \\[0.15cm] 
5546986344820400512 & ($125.63998$, $-32.62$) & $-0.08 \pm 0.022$ & $-1.986 \pm 0.028$ & $2.747 \pm 0.035$ & $79.255 \pm 1.273$ & $13.82$ & $29062^{+5928}_{-4950}$ & $32552^{+5782}_{-4781}$ & $551^{+90}_{-75}$ & $0.00$ & $0.93$ \\[0.15cm]

5257182876777912448 & ($144.73682$, $-60.53137$) & $-0.012 \pm 0.017$ & $-3.736 \pm 0.029$ & $3.444 \pm 0.027$ & $22.64 \pm 1.723$ & $13.49$ & $26140^{+6400}_{-4240}$ & $25824^{+6144}_{-3989}$ & $605^{+148}_{-93}$ & $0.03$ & $0.92$ \\[0.15cm]
4326973843264734208 & ($248.8923$, $-14.51844$) & $0.199 \pm 0.031$ & $-20.546 \pm 0.05$ & $-33.974 \pm 0.033$ & $-220.392 \pm 2.052$ & $13.5$ & $5257^{+881}_{-677}$ & $3842^{+450}_{-465}$ & $766^{+163}_{-122}$ & $0.04$ & $0.91$ \\[0.15cm]

5298599521278293504 & ($140.14259$, $-62.46243$) & $-0.053 \pm 0.02$ & $-2.373 \pm 0.071$ & $3.883 \pm 0.055$ & $54.363 \pm 1.17$ & $13.39$ & $28525^{+6774}_{-5110}$ & $28145^{+6545}_{-4850}$ & $579^{+139}_{-104}$ & $0.03$ & $0.88$ \\[0.15cm] 
6700075834174889472 & ($304.32289$, $-32.41577$) & $0.054 \pm 0.037$ & $-7.243 \pm 0.065$ & $4.955 \pm 0.047$ & $22.491 \pm 2.057$ & $12.76$ & $13068^{+3816}_{-3123}$ & $7584^{+3330}_{-2219}$ & $698^{+152}_{-120}$ & $0.10$ & $0.84$ \\[0.15cm] 
4073247619504712192 & ($280.26863$, $-26.28806$) & $0.05 \pm 0.024$ & $-3.596 \pm 0.046$ & $6.231 \pm 0.039$ & $-191.767 \pm 2.735$ & $13.58$ & $14653^{+4331}_{-2807}$ & $6884^{+4240}_{-2648}$ & $695^{+139}_{-88}$ & $0.11$ & $0.84$ \\[0.15cm]

6492391900301222656 & ($348.64665$, $-58.42957$) & $0.095 \pm 0.018$ & $7.502 \pm 0.027$ & $-15.822 \pm 0.026$ & $-149.856 \pm 1.163$ & $13.36$ & $10276^{+1878}_{-1541}$ & $9641^{+1335}_{-944}$ & $658^{+149}_{-117}$ & $0.06$ & $0.84$ \\[0.15cm] 
4596514892566325504 & ($268.57736$, $29.12348$) & $0.064 \pm 0.013$ & $-1.086 \pm 0.019$ & $-10.512 \pm 0.023$ & $-112.792 \pm 1.093$ & $13.49$ & $14255^{+2485}_{-1839}$ & $12120^{+2106}_{-1453}$ & $617^{+121}_{-90}$ & $0.07$ & $0.84$ \\[0.15cm]
5830109386395388544 & ($249.9792$, $-61.90285$) & $-0.006 \pm 0.019$ & $-1.072 \pm 0.027$ & $3.932 \pm 0.029$ & $143.395 \pm 0.633$ & $13.14$ & $23852^{+6287}_{-4917}$ & $17735^{+6123}_{-4680}$ & $600^{+118}_{-88}$ & $0.08$ & $0.84$ \\[0.15cm] 
1990547230937629696 & ($344.00637$, $53.61551$) & $0.043 \pm 0.017$ & $-4.769 \pm 0.028$ & $-2.83 \pm 0.027$ & $-83.38 \pm 1.158$ & $13.31$ & $17543^{+4372}_{-3415}$ & $21331^{+4114}_{-3130}$ & $563^{+112}_{-84}$ & $0.05$ & $0.83$ \\[0.15cm]
5321157479786017280 & ($128.82063$, $-53.20458$) & $-0.023 \pm 0.018$ & $-2.518 \pm 0.032$ & $3.224 \pm 0.034$ & $81.295 \pm 0.668$ & $13.59$ & $27523^{+6086}_{-5176}$ & $28715^{+5877}_{-4914}$ & $545^{+110}_{-95}$ & $0.08$ & $0.83$ \\[0.15cm]

\hline
\end{tabular}

\begin{tablenotes}

\item \emph{Note.} Distances and total velocities are quoted in terms of the median of the distribution, with uncertainties derived from the $16$th and $84$th percentiles.

\end{tablenotes}

\end{threeparttable}
\end{table}
\end{landscape}

\bibliographystyle{mnras}
\bibliography{hvs.bib}

\appendix

\section{Choice of the Prior Probability on Distances}
\label{appendix:prior}

\begin{figure}
	\centering
	\includegraphics[width=0.5\textwidth]{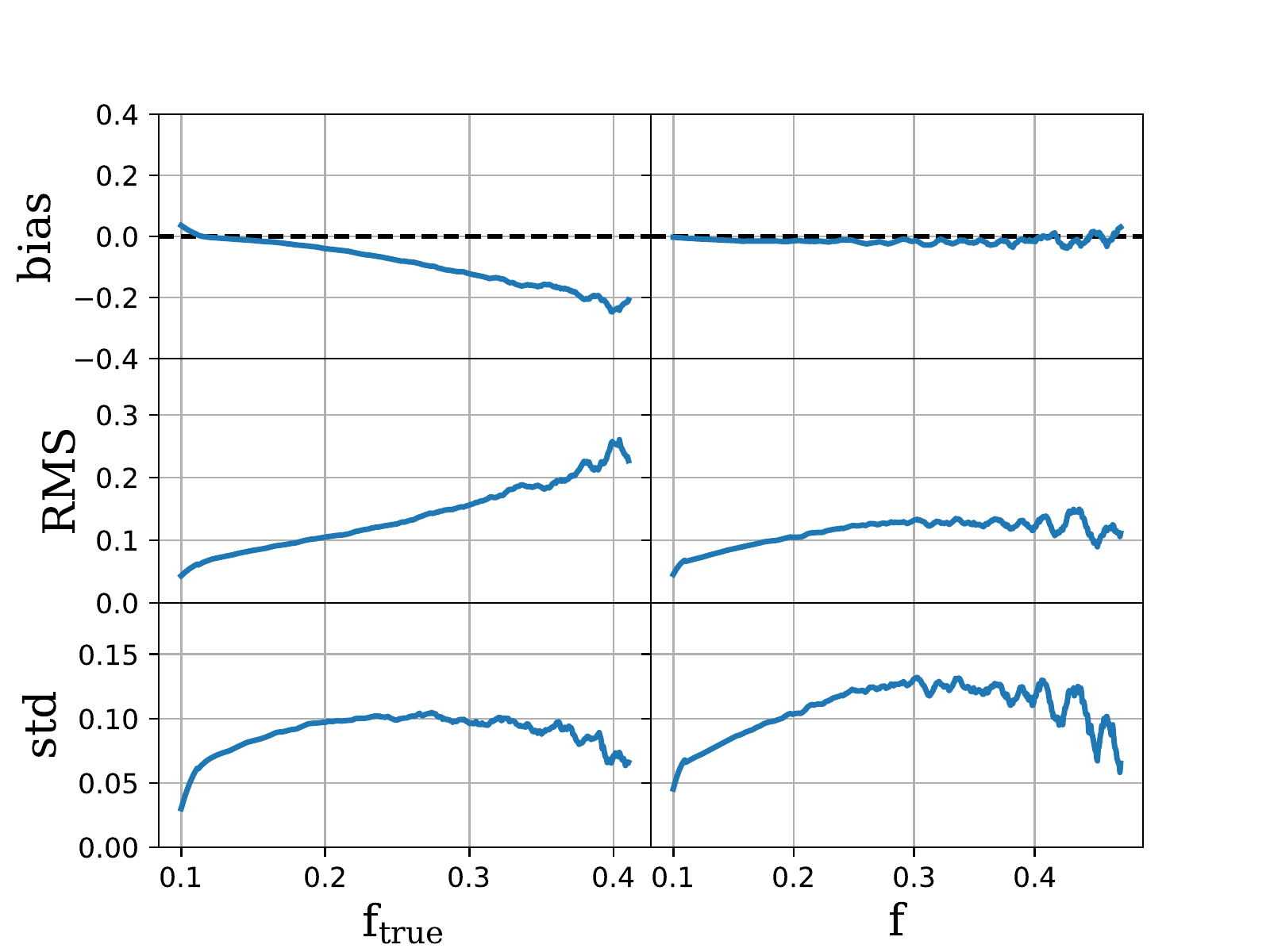}
	\caption{Bias, RMS, and standard deviation of the estimator $x_0$ as a function of $f_\mathrm{true} = \sigma_\varpi d_\mathrm{true}$ (left panel) and $f = \sigma_\varpi/\varpi$ (right panel). The modes of the posterior distributions are estimated using the exponentially decreasing prior with a characteristic scale length $L = 2600$ pc.}
	\label{fig:est_2600}
\end{figure}

In this appendix we discuss the choice of the prior probability on distances $P(d)$ which gives the most accurate results on the subsample of bright stars in \Gaia DR2 with a large relative error on parallax (the \emph{high-f} sample introduced in Section \ref{sec:dist}). We cross-match the \Gaia Universe Model Snapshot \citep[GUMS,][]{robin+12} and the \Gaia Object Generator \citep[GOG,][]{luri+14} catalogues based on the value of the source identifier, to get a resulting sample of $7 \cdot 10^6$ stars with $G_\mathrm{RVS} < 12.2$. We use the latest versions of these mock catalogues, GUMS-18 and GOG-18\footnote{\url{https://wwwhip.obspm.fr/gaiasimu/}}. The resulting combined catalogue contains positions, parallaxes, proper motions, radial velocities, and distances for all stars, with corresponding uncertainties. We extend the limiting magnitude to $G_\mathrm{RVS} = 12.2$ to take into account the fact that \Gaia does take spectra of some stars which are fainter than the limiting magnitude. In particular, these faint stars are the one with the largest error on parallax, so we want to be sure to include them, in order to derive accurate distances for the stars in \Gaia DR2. We multiply the uncertainties on parallax and radial velocity by a factor $(60/22)^{0.5}$, and the ones on both proper motions by a factor $(60/22)^{1.5}$, to simulate the reduced performance of the \Gaia satellite on $22$ months of collected data.

We find $352010$ of the $7$ million stars to have $f = \sigma_\varpi / \varpi > 0.1$. We can see that this value is about 5 times smaller than the one found in \Gaia DR2 (see Section \ref{sec:highf}). All these stars are found at distance larger than $\sim 4.5$ kpc from the Sun, and therefore we choose to adopt the exponentially decreasing prior to derive their distances \citep{astraatmadjaII}, see equation \eqref{eq:prior}. The mode of the posterior distribution in equation \eqref{eq:post} can be determined by numerically finding the roots of the implicit equation \citep{bailer-jones}:
\begin{equation}
\label{eq:implicit}
\frac{d^3}{L} - 2d^2 + \frac{\varpi}{\sigma_\varpi^2}d - \frac{1}{\sigma_\varpi^2} = 0.
\end{equation}
We compute the mode $d_\mathrm{Mo, i}$ for each star $i$ in the simulated catalogue for different values of the scaling length $L$. We then determine the best fitting value of the parameter $L$ as the one minimizing the quantity $\sum_i x_i^2$, where the scaled residual $x_i$ is computed as \citep{astraatmadjaI}:
\begin{equation}
\label{scaled_res}
x_i = \frac{d_\mathrm{Mo, i} - d_\mathrm{true, i}}{d_\mathrm{true, i}},
\end{equation}
where $d_\mathrm{true, i}$ denotes the \emph{true} simulated distance of the $i$-th star. We find the value for the scale length $L = 2600$ pc to work best on this sample of $\sim 352000$ simulated stars. In Fig. \ref{fig:est_2600} we plot the mean value of the bias $\bar{x}$, the root mean squared (RMS) $\bar{x^2}^{1/2}$, and the standard deviation of the residual $x$ for each bin of $f_\mathrm{true} = \sigma_\varpi d_\mathrm{true}$ (left panel) and $f$ (right panel). We can see that, with this choice of prior, the mode of the posterior distribution on distances is an unbiased estimator for all the range of \emph{observed} relative errors in parallax $f$, even if it shows a negative bias of $\sim 20\%$ for stars with large values of the \emph{true} relative error $f_\mathrm{true}$.

The reason why we choose not to use distances from \cite{bailerjones+18} is that the authors fit the values of the scale length $L$ to a full three-dimensional model of the Galaxy\footnote{Note that \cite{bailerjones+18} adopt a scale length that varies smoothly with Galactic longitude and latitude.}. Their values are therefore driven by nearby, bright disk stars, with $f \ll 1$. Such an approach would underestimate distances (and therefore total velocities) to faint distant stars, the ones we are more interested in.

\section{Content of the Distance and Velocity Catalogue}
\label{appendix:catalogue}

\begin{table*}
	
	\caption{Catalogue description. Derived distances and velocities correspond to the median of the distribution, and lower and upper uncertainties are derived, respectively, from the $16$th and $84$th percentiles of the distribution function. Entries labelled $^1$ are derived in this paper, while entries labelled $^2$ are taken from the \Gaia DR2 catalogue \citep{gaiadr2}.}
	
	\label{tab:catalogue}
	
	\begin{threeparttable}
	\setlength{\tabcolsep}{12pt}
		
		\begin{tabular}{llll}
		
			\hline		
			Column & Units & Name & Description \\
			\hline
			
			
			1 & - & source\_id & \Gaia DR2 identifier$^2$ \\
			2 & deg & ra & Right ascension$^2$ \\
			3 & deg & dec & Declination$^2$ \\
			4 & mas & parallax & Parallax$^2$ \\
			5 & mas & e\_parallax & Standard uncertainty in parallax$^2$ \\
			6 & mas yr$^{-1}$ & pmra & Proper motion in right ascension$^2$ \\
			7 & mas yr$^{-1}$ & e\_pmra & Standard uncertainty in proper motion in right ascension$^2$ \\
			8 & mas yr$^{-1}$ & pmdec & Proper motion in declination$^2$ \\
			9 & mas yr$^{-1}$ & e\_pmdec & Standard uncertainty in proper motion declination$^2$ \\
			10 & km s$^{-1}$ & vrad & Radial velocity$^2$ \\
			11 & km s$^{-1}$ & e\_vrad & Radial velocity error$^2$ \\
			12 & mag & GMag & G-band mean magnitude$^2$ \\
			
			
			13 & pc & dist & Distance estimate$^1$ \\
			14 & pc & el\_dist & Lower uncertainty on distance$^1$ \\
			15 & pc & eu\_dist & Upper uncertainty on distance$^1$ \\
			16 & pc & rGC & Spherical Galactocentric radius$^1$ \\
			17 & pc & el\_rGC & Lower uncertainty on spherical Galactocentric radius$^1$ \\
			18 & pc & eu\_rGC & Upper uncertainty on spherical Galactocentric radius$^1$ \\
			19 & pc & RGC & Cylindrical Galactocentric radius$^1$ \\
			20 & pc & el\_RGC & Lower uncertainty on cylindrical Galactocentric radius$^1$ \\
			21 & pc & eu\_RGC & Upper uncertainty on cylindrical Galactocentric radius$^1$ \\
			22 & pc & xGC & Cartesian Galactocentric $x$-coordinate$^1$ \\
			23 & pc & el\_xGC & Lower uncertainty on Cartesian Galactocentric $x$-coordinate$^1$ \\
			24 & pc & eu\_xGC & Upper uncertainty on Cartesian Galactocentric $x$-coordinate$^1$ \\		
			25 & pc & yGC & Cartesian Galactocentric $y$-coordinate$^1$ \\
			26 & pc & el\_yGC & Lower uncertainty on Cartesian Galactocentric $y$-coordinate$^1$ \\
			27 & pc & eu\_yGC & Upper uncertainty on Cartesian Galactocentric $y$-coordinate$^1$ \\
			28 & pc & zGC & Cartesian Galactocentric $z$-coordinate$^1$ \\
			29 & pc & el\_zGC & Lower uncertainty on Cartesian Galactocentric $z$-coordinate$^1$ \\
			30 & pc & eu\_zGC & Upper uncertainty on Cartesian Galactocentric $z$-coordinate$^1$ \\
		
			
			31 & km s$^{-1}$ & U & Cartesian Galactocentric $x$-velocity$^1$ \\
			32 & km s$^{-1}$ & el\_U & Lower uncertainty on Cartesian Galactocentric $x$-velocity$^1$ \\
			33 & km s$^{-1}$ & eu\_U & Upper uncertainty on Cartesian Galactocentric $x$-velocity$^1$ \\
			34 & km s$^{-1}$ & V & Cartesian Galactocentric $y$-velocity$^1$ \\
			35 & km s$^{-1}$ & el\_V & Lower uncertainty on Cartesian Galactocentric $y$-velocity$^1$ \\
			36 & km s$^{-1}$ & eu\_V & Upper uncertainty on Cartesian Galactocentric $y$-velocity$^1$ \\
			37 & km s$^{-1}$ & W & Cartesian Galactocentric $z$-velocity$^1$ \\
			38 & km s$^{-1}$ & el\_W & Lower uncertainty on Cartesian Galactocentric $z$-velocity$^1$ \\
			39 & km s$^{-1}$ & eu\_W & Upper uncertainty on Cartesian Galactocentric $z$-velocity$^1$ \\
			40 & km s$^{-1}$ & UW & Cartesian Galactocentric $xz$-velocity$^1$ \\
			41 & km s$^{-1}$ & el\_UW & Lower uncertainty on Cartesian Galactocentric $xz$-velocity$^1$ \\
			42 & km s$^{-1}$ & eu\_UW & Upper uncertainty on Cartesian Galactocentric $xz$-velocity$^1$ \\ 	
			43 & km s$^{-1}$ & vR & Cylindrical Galactocentric $R$-velocity$^1$ \\
			44 & km s$^{-1}$ & el\_vR & Lower uncertainty on cylindrical Galactocentric $R$-velocity$^1$ \\
			45 & km s$^{-1}$ & eu\_vR & Upper uncertainty on cylindrical Galactocentric $R$-velocity$^1$ \\
			46 & km s$^{-1}$ & vtot & Total velocity in the Galactic rest-frame$^1$ \\
			47 & km s$^{-1}$ & el\_vtot & Lower uncertainty on total velocity in the Galactic rest-frame$^1$ \\
			48 & km s$^{-1}$ & eu\_vtot & Upper uncertainty on total velocity in the Galactic rest-frame$^1$ \\
			
			49 & - & P\_ub & Probability of being unbound from the Galaxy$^1$ \\

			\hline
		
		\end{tabular}
		
	\end{threeparttable}
	
\end{table*}

Table \ref{tab:catalogue} provides an explanation of the content of the catalogue containing distances and velocities for the $7183262$ stars with a radial velocity measurement in \Gaia DR2. The catalogue is publicly available at \url{http://home.strw.leidenuniv.nl/~marchetti/research.html}.

\section{List of High Velocity Stars with $0.5 < P_\mathrm{ub} \leq 0.8$.}
\label{appendix:candidates_0.5}

In Table \ref{tab:candidates_0.5} we present \Gaia identifiers, distances, and total velocities for the $105$ high velocity stars discussed in Section \ref{sec:highV}, with $0.5 < P_\mathrm{ub} \leq 0.8$.

\begin{table}

\caption{Distances and total velocities in the Galactic rest frame for the $105$ ``clean" high velocity star candidates with $0.5 < P_\mathrm{ub} \leq 0.8$. Sources are sorted by decreasing $P_\mathrm{ub}$.}
\label{tab:candidates_0.5}

\begin{threeparttable}

\begin{tabular}{lccc}

\hline
\Gaia DR2 ID & $d$ &  $v_\mathrm{GC}$ & $P_\mathrm{ub}$ \\
 &  (pc) & (\kms) &   \\
\hline

5718618735518384768 & $31308^{+6464}_{-6622}$ & $488^{+69}_{-73}$ & $0.79$ \\[0.15cm] 
4532372476587492608 & $14132^{+3798}_{-2288}$ & $606^{+142}_{-83}$ & $0.78$ \\[0.15cm] 
4366218814874247424 & $7506^{+1521}_{-955}$ & $678^{+137}_{-86}$ & $0.78$ \\[0.15cm] 
5244448023850619648 & $16553^{+3638}_{-2756}$ & $552^{+91}_{-66}$ & $0.77$ \\[0.15cm] 
1994938164981988864 & $22185^{+5526}_{-4751}$ & $516^{+85}_{-70}$ & $0.77$ \\[0.15cm] 
2159020415489897088 & $7686^{+1651}_{-1293}$ & $603^{+123}_{-97}$ & $0.77$ \\[0.15cm] 
2112308930997657728 & $6114^{+999}_{-712}$ & $619^{+119}_{-84}$ & $0.77$ \\[0.15cm] 
5802638672467252736 & $9985^{+1804}_{-1322}$ & $647^{+150}_{-108}$ & $0.76$ \\[0.15cm] 
5996908319666721792 & $13616^{+3593}_{-2595}$ & $662^{+151}_{-108}$ & $0.75$ \\[0.15cm] 
5316722526615701504 & $24242^{+6103}_{-4691}$ & $525^{+123}_{-89}$ & $0.74$ \\[0.15cm] 
2095259117723646208 & $13359^{+2970}_{-2614}$ & $594^{+134}_{-112}$ & $0.73$ \\[0.15cm] 
5839686407534279808 & $7346^{+1033}_{-839}$ & $633^{+112}_{-92}$ & $0.72$ \\[0.15cm] 
1333199496978208128 & $20038^{+4062}_{-3076}$ & $543^{+120}_{-86}$ & $0.72$ \\[0.15cm] 
2089995308886282880 & $13397^{+2700}_{-1874}$ & $573^{+121}_{-81}$ & $0.71$ \\[0.15cm] 
2045752026157687040 & $11799^{+2705}_{-2004}$ & $604^{+144}_{-106}$ & $0.71$ \\[0.15cm] 
6431596947468407552 & $11356^{+2099}_{-1531}$ & $590^{+66}_{-47}$ & $0.71$ \\[0.15cm] 
5247579810921207680 & $27357^{+5878}_{-4547}$ & $499^{+115}_{-85}$ & $0.7$ \\[0.15cm] 
5298494930231856512 & $23913^{+5493}_{-4057}$ & $510^{+119}_{-85}$ & $0.7$ \\[0.15cm] 
2095397827987170816 & $14751^{+2839}_{-2301}$ & $574^{+122}_{-98}$ & $0.7$ \\[0.15cm] 
4656931544705794816 & $24368^{+5597}_{-4637}$ & $514^{+118}_{-95}$ & $0.7$ \\[0.15cm] 
6642234513167197824 & $6836^{+1252}_{-1037}$ & $649^{+117}_{-91}$ & $0.69$ \\[0.15cm] 
5399966178291369728 & $10155^{+2090}_{-1430}$ & $566^{+121}_{-81}$ & $0.69$ \\[0.15cm] 
5374177064347894272 & $6225^{+1109}_{-879}$ & $587^{+97}_{-76}$ & $0.68$ \\[0.15cm] 
2072048770884296704 & $16139^{+3291}_{-2678}$ & $552^{+118}_{-94}$ & $0.68$ \\[0.15cm] 
6116555426949827200 & $7741^{+1164}_{-1011}$ & $628^{+118}_{-102}$ & $0.67$ \\[0.15cm] 
6500989806352727936 & $10407^{+2456}_{-1809}$ & $577^{+128}_{-90}$ & $0.67$ \\[0.15cm] 
5217818333256869376 & $8642^{+1631}_{-1139}$ & $585^{+118}_{-81}$ & $0.67$ \\[0.15cm] 
2106519830479009920 & $8213^{+1326}_{-1065}$ & $570^{+85}_{-67}$ & $0.67$ \\[0.15cm] 
6397497209236655872 & $5802^{+643}_{-487}$ & $587^{+54}_{-41}$ & $0.66$ \\[0.15cm] 
2044224735768501760 & $15167^{+3227}_{-2538}$ & $560^{+124}_{-96}$ & $0.66$ \\[0.15cm] 
5303927273594669056 & $20331^{+5200}_{-3372}$ & $508^{+118}_{-73}$ & $0.66$ \\[0.15cm] 
1966103266381646720 & $28232^{+6210}_{-5780}$ & $474^{+88}_{-76}$ & $0.65$ \\[0.15cm] 
6241406793347941504 & $14098^{+4035}_{-3000}$ & $609^{+139}_{-98}$ & $0.65$ \\[0.15cm] 
5627896072604568960 & $22754^{+5478}_{-4591}$ & $490^{+101}_{-83}$ & $0.65$ \\[0.15cm] 
5415267600583814912 & $24505^{+6046}_{-4520}$ & $498^{+115}_{-87}$ & $0.65$ \\[0.15cm] 
5856098302217892352 & $19735^{+4562}_{-3404}$ & $529^{+127}_{-93}$ & $0.65$ \\[0.15cm] 
6444276683058885248 & $11413^{+3064}_{-2202}$ & $617^{+147}_{-103}$ & $0.65$ \\[0.15cm] 
2094386346009409280 & $14643^{+2968}_{-2007}$ & $549^{+125}_{-82}$ & $0.64$ \\[0.15cm] 
5309766504975294592 & $25956^{+5528}_{-5114}$ & $490^{+106}_{-96}$ & $0.64$ \\[0.15cm] 
3905884598043829504 & $2709^{+385}_{-289}$ & $580^{+115}_{-86}$ & $0.63$ \\[0.15cm] 
2038012426369296128 & $16453^{+4086}_{-3062}$ & $543^{+127}_{-88}$ & $0.63$ \\[0.15cm] 
5317203154946837760 & $18068^{+3537}_{-3079}$ & $510^{+94}_{-80}$ & $0.63$ \\[0.15cm] 
5897201311028035456 & $17717^{+4423}_{-4116}$ & $543^{+83}_{-70}$ & $0.62$ \\[0.15cm]

\hline
\end{tabular}


\end{threeparttable}
\end{table} 
\begin{table}

\contcaption{.}
\label{tab:candidates_0.5}

\begin{threeparttable}

\begin{tabular}{lccc}

\hline
\Gaia DR2 ID & $d$ &  $v_\mathrm{GC}$ & $P_\mathrm{ub}$ \\
 &  (pc) & (\kms) &   \\
\hline

5823425661366917376 & $15652^{+4759}_{-3695}$ & $568^{+127}_{-97}$ & $0.62$ \\[0.15cm] 
5807202126764572288 & $14365^{+3602}_{-2776}$ & $563^{+97}_{-74}$ & $0.62$ \\[0.15cm] 
3705761936916676864 & $3756^{+371}_{-300}$ & $566^{+59}_{-46}$ & $0.62$ \\[0.15cm] 
2183775885439262592 & $23213^{+5580}_{-4338}$ & $480^{+102}_{-78}$ & $0.62$ \\[0.15cm] 
5317776481532378240 & $19139^{+4400}_{-3115}$ & $500^{+112}_{-79}$ & $0.62$ \\[0.15cm] 
6077622510498751616 & $14503^{+3852}_{-2502}$ & $538^{+84}_{-46}$ & $0.62$ \\[0.15cm] 
4531575708618805376 & $12030^{+2748}_{-1974}$ & $562^{+80}_{-56}$ & $0.62$ \\[0.15cm] 
1956680279930601344 & $23550^{+6723}_{-4451}$ & $480^{+113}_{-75}$ & $0.62$ \\[0.15cm] 
6010197124582216832 & $10863^{+3441}_{-1945}$ & $629^{+118}_{-65}$ & $0.62$ \\[0.15cm] 
5232568213032618496 & $27921^{+5690}_{-4842}$ & $487^{+111}_{-92}$ & $0.61$ \\[0.15cm] 
5249820306388948992 & $26092^{+6478}_{-4213}$ & $478^{+117}_{-78}$ & $0.61$ \\[0.15cm] 
5779439836114210304 & $23901^{+5743}_{-4509}$ & $492^{+69}_{-53}$ & $0.6$ \\[0.15cm] 
5247264629041172608 & $20274^{+3940}_{-3336}$ & $507^{+100}_{-80}$ & $0.6$ \\[0.15cm] 
5912922197004254848 & $12401^{+3128}_{-2696}$ & $610^{+122}_{-99}$ & $0.6$ \\[0.15cm] 
5247811567357582336 & $21321^{+4641}_{-3453}$ & $497^{+114}_{-86}$ & $0.59$ \\[0.15cm] 
4489509905555953408 & $11610^{+2734}_{-2257}$ & $590^{+117}_{-91}$ & $0.59$ \\[0.15cm] 
2121857472227927168 & $13251^{+2401}_{-1679}$ & $522^{+92}_{-63}$ & $0.59$ \\[0.15cm] 
1989862986804105344 & $10429^{+2057}_{-1607}$ & $523^{+107}_{-82}$ & $0.58$ \\[0.15cm] 
6677910160794903296 & $4345^{+554}_{-396}$ & $604^{+106}_{-76}$ & $0.58$ \\[0.15cm] 
6229070238523155328 & $13987^{+4361}_{-2810}$ & $567^{+142}_{-87}$ & $0.58$ \\[0.15cm] 
4452929978332889216 & $24168^{+5324}_{-4537}$ & $496^{+108}_{-88}$ & $0.58$ \\[0.15cm] 
5785402796909679744 & $14723^{+3134}_{-2187}$ & $543^{+132}_{-89}$ & $0.58$ \\[0.15cm] 
5362114562797004544 & $23461^{+5342}_{-4015}$ & $479^{+113}_{-80}$ & $0.57$ \\[0.15cm] 
1331585993728475264 & $10902^{+2413}_{-1920}$ & $544^{+115}_{-87}$ & $0.57$ \\[0.15cm] 
6733156428223193856 & $13978^{+3684}_{-2829}$ & $601^{+122}_{-92}$ & $0.57$ \\[0.15cm] 
6221350429945324032 & $8878^{+2117}_{-1582}$ & $593^{+141}_{-104}$ & $0.57$ \\[0.15cm] 
3454083549225619712 & $5943^{+794}_{-627}$ & $522^{+100}_{-77}$ & $0.57$ \\[0.15cm] 
6868478546915992320 & $14043^{+4460}_{-3582}$ & $576^{+130}_{-101}$ & $0.57$ \\[0.15cm] 
4127621699294858368 & $13174^{+3602}_{-2904}$ & $615^{+128}_{-98}$ & $0.56$ \\[0.15cm] 
1364548016594914560 & $10327^{+1989}_{-1642}$ & $531^{+66}_{-50}$ & $0.56$ \\[0.15cm] 
4609875745549298688 & $10640^{+1380}_{-1204}$ & $544^{+76}_{-66}$ & $0.56$ \\[0.15cm] 
5212817273334550016 & $3811^{+330}_{-283}$ & $565^{+59}_{-51}$ & $0.56$ \\[0.15cm] 
1268023196461923712 & $4586^{+500}_{-390}$ & $568^{+79}_{-61}$ & $0.56$ \\[0.15cm] 
1696697285206197248 & $23235^{+5014}_{-3909}$ & $464^{+111}_{-81}$ & $0.56$ \\[0.15cm] 
6034352158118691072 & $11013^{+2964}_{-2267}$ & $646^{+157}_{-104}$ & $0.56$ \\[0.15cm] 
2098831980759357696 & $15685^{+3439}_{-2694}$ & $518^{+119}_{-92}$ & $0.56$ \\[0.15cm] 
5354094037807264384 & $11683^{+2120}_{-1758}$ & $533^{+111}_{-90}$ & $0.56$ \\[0.15cm] 
4220617568115374848 & $4978^{+814}_{-677}$ & $603^{+114}_{-92}$ & $0.56$ \\[0.15cm] 
5779919841659989120 & $10641^{+2101}_{-1505}$ & $568^{+135}_{-95}$ & $0.55$ \\[0.15cm] 
5317675979297751040 & $27098^{+5311}_{-4561}$ & $451^{+81}_{-70}$ & $0.55$ \\[0.15cm] 
3891412241883772928 & $7004^{+1531}_{-1150}$ & $539^{+88}_{-65}$ & $0.55$ \\[0.15cm] 
4916199478888664320 & $5579^{+725}_{-629}$ & $549^{+66}_{-56}$ & $0.55$ \\[0.15cm] 
2255126837089768192 & $24623^{+4714}_{-4286}$ & $456^{+85}_{-74}$ & $0.55$ \\[0.15cm] 
5511130239834500864 & $20579^{+5603}_{-3668}$ & $467^{+100}_{-68}$ & $0.55$ \\[0.15cm]

\hline
\end{tabular}


\end{threeparttable}
\end{table} 
\begin{table}

\contcaption{.}
\label{tab:candidates_0.5}

\begin{threeparttable}

\begin{tabular}{lccc}

\hline
\Gaia DR2 ID & $d$ &  $v_\mathrm{GC}$ & $P_\mathrm{ub}$ \\
 &  (pc) & (\kms) &   \\
\hline

3784964943489710592 & $4031^{+733}_{-505}$ & $552^{+92}_{-61}$ & $0.55$ \\[0.15cm] 
2038818952503671424 & $26358^{+5535}_{-5090}$ & $469^{+105}_{-92}$ & $0.55$ \\[0.15cm] 
1954400884950622464 & $19455^{+4960}_{-3498}$ & $482^{+108}_{-75}$ & $0.54$ \\[0.15cm] 
5846560382443820032 & $7054^{+936}_{-629}$ & $585^{+96}_{-64}$ & $0.54$ \\[0.15cm] 
6130863887159694848 & $9639^{+2070}_{-1335}$ & $550^{+133}_{-85}$ & $0.54$ \\[0.15cm] 
5231000034569444992 & $18206^{+3340}_{-3437}$ & $501^{+101}_{-105}$ & $0.53$ \\[0.15cm] 
2186887606421426816 & $24376^{+4607}_{-4292}$ & $454^{+74}_{-67}$ & $0.53$ \\[0.15cm] 
5818738237122521344 & $11884^{+3059}_{-2216}$ & $559^{+136}_{-89}$ & $0.53$ \\[0.15cm] 
5249917441371959040 & $17540^{+4063}_{-3149}$ & $494^{+116}_{-85}$ & $0.53$ \\[0.15cm] 
6639557580310606976 & $11135^{+3975}_{-2226}$ & $579^{+108}_{-55}$ & $0.53$ \\[0.15cm] 
4210389120686616832 & $7886^{+2550}_{-1822}$ & $599^{+143}_{-88}$ & $0.52$ \\[0.15cm] 
1191989287342960640 & $10798^{+2233}_{-1691}$ & $549^{+131}_{-96}$ & $0.52$ \\[0.15cm] 
6098331056080412416 & $16089^{+3894}_{-3358}$ & $528^{+89}_{-72}$ & $0.52$ \\[0.15cm] 
2086507417487662976 & $26304^{+5278}_{-4208}$ & $448^{+90}_{-72}$ & $0.51$ \\[0.15cm] 
5303240216263896192 & $21972^{+5482}_{-3995}$ & $464^{+111}_{-79}$ & $0.51$ \\[0.15cm] 
2000253135474943616 & $16537^{+3984}_{-3129}$ & $475^{+89}_{-69}$ & $0.51$ \\[0.15cm] 
6035120957243593600 & $10873^{+3525}_{-2307}$ & $603^{+124}_{-76}$ & $0.51$ \\[0.15cm] 
1612628419987892096 & $25402^{+5063}_{-3992}$ & $442^{+104}_{-79}$ & $0.5$ \\[0.15cm] 
 
\hline
\end{tabular}


\end{threeparttable}
\end{table}

\section{Global Parallax Offset}
\label{appendix:offset}

In this appendix we discuss the impact of including the $-0.029$ mas global parallax zeropoint mentioned in \cite{lindegren+18}, derived from \Gaia's observations of distant quasars. Being a negative offset, the net effect is to lower the inferred distances, and therefore the resulting total velocities. We repeat the Bayesian analysis discussed in Section \ref{sec:dist} to the $20$ stars with $P_\mathrm{ub} > 80 \% $. In this case, the likelihood probability is again a multivariate Gaussian distribution, but with mean vector \citep{bailerjones+18}:
\begin{equation}
\label{eq:mean_zp}
\mathbf{m} = [\mu_{\alpha*}, \mu_\delta, 1/d + \varpi_\mathrm{zp}],
\end{equation}
where $\varpi_\mathrm{zp} = -0.029$ mas. In Table \ref{tab:candidates_zp} we report the updated values of the distance, total velocity, and probability of being unbound from the Galaxy for the $20$ stars discussed in Section \ref{sec:highVcand}. We now find 14 candidates ($70\%$) to have an updated $P_\mathrm{ub} > 50\%$, and $4$ stars ($20\%$) to have $P_\mathrm{ub} > 80\%$.

\begin{table}

\caption{Distances and total velocities in the Galactic rest frame for the $20$ ``clean" high velocity star candidates with $P_\mathrm{ub} > 0.8$ presented in Table \ref{tab:candidates}, including the $-0.029$ mas global parallax offset. For comparison, stars are sorted as in Table \ref{tab:candidates}.}
\label{tab:candidates_zp}

\begin{threeparttable}

\begin{tabular}{lccc}

\hline
\Gaia DR2 ID & $d$ &  $v_\mathrm{GC}$ & $P_\mathrm{ub}$ \\
 &  (pc) & (\kms) &   \\
\hline

5932173855446728064 & $2096^{+130}_{-117}$ & $747^{+3}_{-3}$ & $1.0$ \\[0.15cm] 
1383279090527227264 & $7144^{+809}_{-782}$ & $745^{+105}_{-102}$ & $0.98$ \\[0.15cm]
6456587609813249536 & $7964^{+1297}_{-885}$ & $660^{+135}_{-92}$ & $0.82$ \\[0.15cm]
5935868592404029184 & $10010^{+2144}_{-1800}$ & $665^{+81}_{-67}$ & $0.75$ \\[0.15cm] 
5831614858352694400 & $17160^{+4736}_{-4055}$ & $600^{+101}_{-86}$ & $0.73$ \\[0.15cm] 
5239334504523094784 & $14426^{+3339}_{-2236}$ & $454^{+105}_{-66}$ & $0.32$ \\[0.15cm] 
4395399303719163904 & $9934^{+2389}_{-1586}$ & $535^{+112}_{-71}$ & $0.37$ \\[0.15cm] 
1396963577886583296 & $23038^{+5341}_{-3347}$ & $511^{+112}_{-68}$ & $0.73$ \\[0.15cm]
5593107043671135744 & $32604^{+6740}_{-4982}$ & $511^{+79}_{-61}$ & $0.9$ \\[0.15cm] 
5546986344820400512 & $26048^{+6507}_{-4962}$ & $507^{+99}_{-74}$ & $0.78$ \\[0.15cm] 
5257182876777912448 & $21973^{+4863}_{-4292}$ & $515^{+106}_{-91}$ & $0.66$ \\[0.15cm] 
4326973843264734208 & $4718^{+725}_{-580}$ & $670^{+131}_{-104}$ & $0.72$ \\[0.15cm] 
5298599521278293504 & $24102^{+6820}_{-3800}$ & $489^{+140}_{-74}$ & $0.63$ \\[0.15cm] 
6700075834174889472 & $11382^{+4021}_{-2622}$ & $631^{+158}_{-98}$ & $0.69$ \\[0.15cm] 
4073247619504712192 & $11656^{+3234}_{-1949}$ & $601^{+101}_{-61}$ & $0.47$ \\[0.15cm] 
6492391900301222656 & $7999^{+1457}_{-1042}$ & $487^{+109}_{-73}$ & $0.29$ \\[0.15cm] 
4596514892566325504 & $10522^{+1717}_{-1145}$ & $436^{+83}_{-53}$ & $0.14$ \\[0.15cm] 
5830109386395388544 & $19057^{+4550}_{-3307}$ & $514^{+84}_{-59}$ & $0.51$ \\[0.15cm] 
1990547230937629696 & $13243^{+2851}_{-2563}$ & $456^{+71}_{-62}$ & $0.37$ \\[0.15cm] 
5321157479786017280 & $22613^{+5272}_{-4543}$ & $456^{+94}_{-81}$ & $0.51$ \\[0.15cm]

\hline
\end{tabular}




\end{threeparttable}
\end{table}

\section{Systematic Errors in Parallax}
\label{appendix:inflated_erros}

\Gaia DR2 uncertainties in parallax do not include the contribution from \emph{systematic} errors, which might depend on the magnitude, position, colour, and other property of the source. The mean value of the systematic errors is the global parallax offset $\varpi_\mathrm{zp}$ already discussed in Appendix \ref{appendix:offset}. In this appendix we discuss the impact of adding this contribution to the quoted values of the parallax uncertainties. To do that, we follow the advice and guidelines presented in \cite{Lindegren+18p}. Internal uncertainties published in the \Gaia DR2 catalogue can be artificially inflated to keep into account systematic errors \citep[e.g.][]{lindegren+16}:
\begin{equation}
\label{eq:inflated_errors}
\sigma_{\varpi, \mathrm{ext}} = \sqrt{ k^2 \sigma_\varpi + \sigma_s^2},
\end{equation}
where $k \gtrsim$ 1 is a correction factor, and $\sigma_s$ is the variance of the systematic error. These parameters need to be calibrated using external datasets. \cite{Lindegren+18p} suggest adopting $k = 1.08$, $\sigma_s = 0.021$ mas ($k = 1.08$, $\sigma_s = 0.043$ mas) for bright stars with $G \lesssim 13$ (faint stars with $G \gtrsim 13$). In Table \ref{tab:candidates_inflated_erros} we report the updated values for distances, total velocities, and probability of being unbound from the Galaxy for the sample of $20$ stars discusses in Section \ref{sec:highVcand}. All of the stars but one are classified as \emph{faint} stars. $9$ ($5$) stars out of $20$ now have an updated probability $P_\mathrm{ub} > 0.5$ ($P_\mathrm{ub} > 0.8$). We want to stress that the adopted value for $\sigma_s$ is likely overestimated for the typical magnitude of stars in our sample \citep{Lindegren+18p}, therefore this is a conservative approach, which underestimates distances (and therefore total velocities).

\begin{table}

\caption{Distances and total velocities in the Galactic rest frame for the $20$ ``clean" high velocity star candidates with $P_\mathrm{ub} > 0.8$ presented in Table \ref{tab:candidates}. Parallax uncertainties are inflated according to equation \eqref{eq:inflated_errors}. For comparison, stars are sorted as in Table \ref{tab:candidates}.}
\label{tab:candidates_inflated_erros}

\begin{threeparttable}

\begin{tabular}{lccc}

\hline
\Gaia DR2 ID & $d$ &  $v_\mathrm{GC}$ & $P_\mathrm{ub}$ \\
 &  (pc) & (\kms) &   \\
\hline

5932173855446728064 & $2316^{+306}_{-265}$ & $746^{+3}_{-3}$ & $1.0$ \\[0.15cm] 
1383279090527227264 & $8577^{+3716}_{-2135}$ & $931^{+484}_{-278}$ & $0.94$ \\[0.15cm] 
6456587609813249536 & $9370^{+3917}_{-2262}$ & $806^{+414}_{-234}$ & $0.86$ \\[0.15cm] 
5935868592404029184 & $10744^{+3489}_{-2685}$ & $694^{+131}_{-101}$ & $0.8$ \\[0.15cm] 
5831614858352694400 & $13924^{+5147}_{-3860}$ & $531^{+109}_{-79}$ & $0.53$ \\[0.15cm] 
5239334504523094784 & $12051^{+4102}_{-3115}$ & $384^{+124}_{-84}$ & $0.22$ \\[0.15cm] 
4395399303719163904 & $11019^{+3704}_{-3061}$ & $585^{+179}_{-138}$ & $0.58$ \\[0.15cm] 
1396963577886583296 & $15707^{+5086}_{-3944}$ & $372^{+92}_{-54}$ & $0.21$ \\[0.15cm] 
5593107043671135744 & $18643^{+5317}_{-4575}$ & $348^{+62}_{-53}$ & $0.14$ \\[0.15cm] 
5546986344820400512 & $16803^{+5056}_{-4307}$ & $371^{+77}_{-61}$ & $0.2$ \\[0.15cm] 
5257182876777912448 & $14545^{+4481}_{-3702}$ & $361^{+91}_{-65}$ & $0.17$ \\[0.15cm] 
4326973843264734208 & $6032^{+2296}_{-1452}$ & $909^{+426}_{-265}$ & $0.91$ \\[0.15cm] 
5298599521278293504 & $16316^{+5884}_{-4573}$ & $341^{+109}_{-77}$ & $0.18$ \\[0.15cm] 
6700075834174889472 & $12278^{+4717}_{-3393}$ & $667^{+187}_{-129}$ & $0.75$ \\[0.15cm] 
4073247619504712192 & $11462^{+4236}_{-2678}$ & $593^{+135}_{-83}$ & $0.48$ \\[0.15cm] 
6492391900301222656 & $9897^{+4543}_{-2488}$ & $630^{+359}_{-185}$ & $0.69$ \\[0.15cm] 
4596514892566325504 & $11421^{+4949}_{-2998}$ & $479^{+242}_{-141}$ & $0.44$ \\[0.15cm] 
5830109386395388544 & $14312^{+5392}_{-4022}$ & $430^{+97}_{-69}$ & $0.27$ \\[0.15cm] 
1990547230937629696 & $11614^{+4698}_{-2769}$ & $416^{+116}_{-66}$ & $0.35$ \\[0.15cm] 
5321157479786017280 & $15167^{+5524}_{-3904}$ & $328^{+95}_{-62}$ & $0.13$ \\[0.15cm]

\hline
\end{tabular}




\end{threeparttable}
\end{table}

\bsp

\label{lastpage}

\end{document}